%% file: main.tex
\documentclass[twocolumn]{aastex63}
\usepackage[percent]{overpic}
\usepackage{amsmath}

\graphicspath{{./}{Figures/}}
\input{med_vmpdlas_defs_v9_noESI}
\turnoffonetrue

\submitjournal{ApJ}

\shorttitle{Empirical Constraints on CCSN Yields}
\shortauthors{Nu\~{n}ez et al.}

\begin{document}

\title{Empirical Constraints on Core Collapse Supernova Yields using Very Metal Poor Damped Lyman Alpha Absorbers}

\correspondingauthor{Evan Haze Nu\~{n}ez}
\email{enunez@astro.caltech.edu}

\author[0000-0001-5595-757X]{Evan H. Nu\~{n}ez}
\affil{California Institute of Technology, 1200 E. California Blvd., MC 249-17, Pasadena, CA 91125, USA}

\author[0000-0001-6196-5162]{Evan N. Kirby}
\affil{California Institute of Technology, 1200 E. California Blvd., MC 249-17, Pasadena, CA 91125, USA}
\affil{Department of Physics, University of Notre Dame, Notre Dame, IN 46556, USA}

\author[0000-0002-4834-7260]{Charles C. Steidel}
\affil{California Institute of Technology, 1200 E. California Blvd., MC 249-17, Pasadena, CA 91125, USA}

\begin{abstract}
We place empirical constraints on the yields from zero- and low-metallicity core collapse supernovae (CCSNe) using abundances measured in very metal-poor (VMP; [Fe/H] $\leq$ $-2$) Damped Lyman Alpha Absorbers (DLAs). For some abundance ratios \edit1{([N,Al,S/Fe])}, VMP DLAs constrain the metal yields of the first SNe more reliably than VMP stars. We compile a large sample of high-S/N VMP DLAs from over 30 years of literature, most with high resolution spectral measurements. We infer the IMF-averaged CCSNe yield from the median values from the DLA abundance ratios of C, N, O, Al, Si, S, and Fe (over Fe and O). We assume that the DLAs are metal-poor enough that they represent galaxies in their earliest stages of evolution, when CCSNe are the only nucleosynthetic sources of the metals we analyze. We compare five sets of zero- and low-metallicity theoretical yields to the empirical yields derived in this work. We find that the five models agree with the DLA yields for ratios containing Si and S. Only one model, \citet[][hereafter HW10]{heger+2010}, reproduced the DLA values for N, and one other model, \citet[][hereafter LC18]{limongi+2018}, reproduced [N/O]. We found little change in the theoretical yields with the adoption of a SN explosion landscape (where certain progenitor masses collapse into black holes, contributing no yields) onto HW10, but fixing explosion energy to progenitor mass results in wide disagreements between the predictions and DLA abundances. We investigate the adoption of a simple, observationally motivated Initial Distribution of Rotational Velocities for LC18 and find a slight improvement.
\end{abstract}


\section{Introduction} \label{sec:intro}
Core Collapse Supernovae (CCSNe) play a vital role in the evolution of the universe. They drive, and/or substantially contribute to, many astrophysical processes including the creation of metals, the dispersal of enriched gas, the injection of large amounts of energy into the interstellar medium, the chemical evolution of galaxies, and the launching of galactic outflows \citep{woosley+2002, pettini_2011}. In other words, CCSNe connect the largest and smallest scales of the universe.

The nearby universe no longer has any CCSNe from zero-metallicity or extremely metal-poor progenitor stars because the timescales associated with CCSNe are much shorter than the timescale for galactic chemical evolution.  Although there is debate on the initial masses---and hence lifetimes---of Population III (PopIII) stars \citep[e.g.,][]{greif+2011,stacy+2016} we have yet to observe a truly metal-free star.  Most PopIII candidates have turned out to be extremely metal-poor PopII stars \citep[e.g.,][]{aguado+2018}. %The mass ranges for Population III stars have been hypothesized to be very massive (\textbf{NUMBER}) and therefore had very short lifetimes ($< 10$ Myr) meaning it is unlikely that any of them have survived until today (\textbf{REFERENCES}). 
The indirect way to estimate yields of low-metallicity CCSNe is galactic archaeology, \textbf{\textnormal{i.e.}} by observing the abundances of metal-poor stars found in pristine environments that condensed from the gas enriched by these CCSNe. Such observations can test theoretical models of CCSN nucleosynthesis.

Modeling the yields of CCSNe is a challenge that is fraught with many uncertainties \citep{woosley+1995, heger+2010, nomoto+2006,romano+2010}. These uncertainties compound with one another in ways that can drastically affect the resulting yield prediction. Uncertainties are introduced in, but are not limited to, the pre-supernova evolution of the star, the adopted nuclear reaction rates, the explosion mechanism, the rotational velocity of the progenitor, and the explosion energy \citep{woosley+1995, nomoto+2006, romano+2010, limongi+2018}. Because of these uncertainties, observational data are sometimes integrated into models, e.g., light curves observed in nearby SNe, the compact remnants of those SNe, and the abundances observed in metal-poor stars \citep[e.g.,][]{woosley+2002,perego+2015}.
%One can see how these can compound since the nuclear reaction rates would sensitively affect the pre-supernova evolution of the star as well as the explosive nucleosynthesis, the pre-supernova evolution of the star fundamentally changes the structure of the star which then affects the explosion mechanism \citep[i.e., mass piston or thermal bomb][]{woosley+1995}. Additionally, the theory that is used to describe a physical process could affect the results such as what theory of convection is adopted \citep[i.e., Schwarzschild or Ledoux criterion][]{woosley+1995, woosley+2002, romano+2010}. All of these effects combine to either skew the resulting modeled yields to either under-represent or over-represent that one would observes in nature.

%One application that CCSNe yields play a critical role in is the modeling galactic chemical evolution (GCE). These models track the abundances of elements over the history of a galaxy. These models can take many forms, from the physics rich simulations \citet[i.e., FIRE-2][]{hopkins+2018} to simpler one-zone models \citet[i.e., OMEGA][]{cote+2016}. Even though the models can range tremendously in their physical diversity, they are fundamentally rely on theoretical stellar yields to inform how CCSNe would enrich the gas reservoir. As mentioned previously, there can be tremendous diversity between different theoretical yields which is discussed by \citet[][R10]{romano+2010}.

Galactic archaeology can help estimate CCSN yields.  Metal-poor stars in the galactic halo and in dwarf galaxies are PopII stars that condensed from gas that was near primordial and therefore likely to be primarily enriched by nucleosynthetic events with short delays, like CCSNe. Their photospheric abundances may reflect the yields of PopIII CCSNe. This method has been used to place empirical constraints on CCSNe.  For example, \citet{kirby+2019}, \cite{delosreyes+2020}, and \citet[][]{ishigaki+2021} used the abundances of metal poor stars in Local Group dwarf galaxies to estimate the yields of both CCSNe and Type Ia SNe.

However, galactic archaeology has some drawbacks.  The abundances measured from stellar atmospheres can depend critically on assumptions that are adopted to interpret their spectra (e.g., LTE).  Some elements are more sensitive to this assumption than others.  For example, non-LTE \textbf{\textnormal{(nLTE)}} corrections to Al abundances can be as large as \textbf{\textnormal{+1~dex}} \citep[][]{nordlander+2017}\edit1{, and 3D nLTE corrections for O and C can be as large as \textbf{\textnormal{-0.6 dex}} and \textbf{\textnormal{-0.3 dex}} respectively \citep[][]{amarsi+2019b}}. Stellar evolution can also alter the atmospheric abundances of some elements.  For example, dredge-up on the red giant branch can deplete C and enhance N\@.  In this way, stars act more as a middle-man to the ``true'' abundances from the near primordial gas out of which they formed. Our aim in this work is to measure this primordial gas directly to avoid the complications and corrections required to accurately map from the stellar spectra to atomic abundance.

Very Metal-Poor (VMP) Damped Lyman Alpha Absorbers (DLA) offer a unique way to investigate nearly primordial gas without the need for stellar spectroscopy. VMP DLAs are effectively the gas from which PopII stars formed.  Therefore, measuring DLA abundances is the same as measuring the near-primordial gas. DLAs are QSO absorbers classified by their large column density $N_{\rm H} > 2 \times 10^{20} \; \rm{cm^{-2}}$ and therefore have damping wings in their Lyman-alpha absorption. %, as example of which is shown in Figure \ref{fig:vmp_dla_ex}. 
\edit1{DLAs account for up to $\sim$90\% of the neutral hydrogen content of the universe at their redshift \citep{lanzetta+1995,prochaska+2009,zafar+2013,sanchez-ramirez+2016}}. 

VMP DLAs at high redshift ($z \geq 2$) have a chemical enrichment history dominated by CCSNe \citep[][]{cooke+2017}. \edit1{This can be seen in their [$\alpha$/Fe] evolution with respect to [Fe/H], which exhibits a knee at ${\rm [Fe/H]} \sim -2$, preceded by a plateau \citep[][]{prochaska+2002,rafelski+2012,berg+2015,cooke+2015}}. This is qualitatively similar to evolution seen in dwarf spheroidal galaxies, where the plateau is indicative of CCSNe dominating the chemical evolution because they produce both $\alpha$ and Fe-peak elements, then Type Ia supernovae (SNe Ia) dominating after the knee because they primarily produce Fe-peak elements \textbf{\textnormal{\citep[e.g.,][]{cooke+2015, berg+2015,skuladottir+2018}}}. \citet{cooke+2015} and \citet{welsh+2019} also showed that winds from AGB stars do not significantly affect the enrichment of VMP DLAs for almost any elements except carbon.
%Type Ia SNe have delay times upwards of 1 Gyr (\textbf{REFERENCES}). Winds from AGB stars have shorter timescales but would only affect certain elements (\textbf{REFERENCES}).

%VMP DLAs are characterized by a metallicity ${\rm [Fe/H]} \leq -2$. This low metallicity points to a simple enrichment history where only one, or a few, bursts of star formation have occurred (\textbf{REFERENCES}). This, coupled with their high redshift, imply that the few enrichment's from their star formation history must be dominated by the lowest metallicity CCSNe. Therefore, the abundances measured in VMP DLAs should be reflective of the yields from these zero metallicity CCSNe.

Deriving abundances from VMP DLAs is relatively straightforward compared to stars. One can map directly from line strength to abundance for most elements of interest. This is primarily due to the fact that lines used to measure metal abundances in VMP DLAs (excluding hydrogen and very strong metal lines, e.g., \ion{O}{1} and \ion{C}{2}) are weak at almost all wavelengths, such that line strength is proportional to column density and hence abundance.  Even so, multiple ionization states and dust depletion (both discussed below) could lead to systematic errors in measuring DLA abundances.  %However, both effects have been shown to be minimal at very low metallicities.

Multiple ionization states of elements in DLAs could lead to overestimating, or underestimating, the ``true'' abundance of a system depending on whether the transition measured is the dominant ionization state. This leads to a discrepancy between the ``true'' abundance and what is measured, which must be corrected when converting to abundance. \edit1{In general, the ionization corrections for DLAs are low due to their high \ion{H}{1} column density, which allows the gas to self-shield from the UV background emitted from quasars and galaxies \citep{prochaska+2002b, zheng+2002, wolfe+2005, cooke+2011_sources, cooke+2016b}.} %Additionally, it has been shown that there is less need for ionization corrections at very low metallicities \citep{cooke+2011_sources,cooke+2017}. 
%Therefore, multiple ionization states are not a significant source of uncertainty for VMP DLAs because there is usually one ionization state that dominates.
\textbf{\textnormal{Therefore ionization corrections are not a significant source of uncertainty in VMP DLA abundances because there is usually only one dominant ionization state detected.}}

\edit1{Dust depletion can lead to an underestimate of DLA abundance \citep{wolfe+2005, berg+2015, decia+2016}}. Refractory elements (e.g., Si, Mg, Fe) could condense onto dust grains, removing their signature from the detectable gas phase which results in an underestimate of the ``true'' abundance. It has been shown that DLAs in the VMP regime need little to no dust corrections \citep{wolfe+2005,pettini_2011, cooke+2011_sources, cooke+2017}. Therefore, dust depletion is not a significant source of uncertainty in VMP DLAs.  Retiring the major possible sources of systematic errors in DLAs makes them superior to PopII stars as sites to examine early nucleosynthesis.

%\textbf{As a summary,} VMP DLAs are ideal for the following reasons 1) Their metal-poor status mean that they have seen only few enrichment events in their history, 2) since they lie at such high redshift, their enrichment would have to have come from primarily CCSNe since there would have been enough time for delayed enriching events to have occurred (i.e., Type Ia  supernovae (SNe Ia) or winds from AGB stars), and 3) measuring abundances from their cool, diffuse gas is straightforward compared to the complications that arise from a stellar atmosphere.

The goal of this paper is to use the abundances measured from VMP DLAs to place empirical constraints on the yields of zero- or low-metallicity CCSNe. %We achieve this by finding the median of their abundance ratios across -4 $<$ [Fe/H] $<$ -2. 
With the empirical constraints in hand, we attempt to quantify the differences between the empirical yields and the most widely adopted zero- and low-metallicity theoretical yields to offer insights into the most important input physics in the various models. %and their evolution through a GCE model. 

This paper is organized as follows. In Section \ref{sec:data} we discuss the sample of VMP DLAs. In Section \ref{sec:empricial_yields} we place empirical constraints on the abundance ratios of zero- to low-metallicity CCSNe and compare them to metal poor stars. In Section \ref{sec:theory_comparison} we compare theoretical yields of zero- and low-metallicty CCSNe to our empirical estimates, taking realistic explosion physics into account and reducing free parameters in a couple of the models in Section \ref{sec:compare_explosion}. In Section \ref{sec:discussion} we discuss the different input physics in the various theoretical yields that seem to reproduce our empirical yields. We summarize our results in Section \ref{sec:conclusions}.
%I convert the constraints of the ratios to constraints on the elemental yields, then compare the constraints to theoretical yields and a galactic chemical evolution model. In Section \ref{sec:future} I discuss next steps for the project.

\section{Data} \label{sec:data}
%\subsection{Sample Selection} \label{sec:sample_selection}
%\subsection{Final Sample} \label{sec:final_sample}
We compiled a large sample (79 total) of VMP DLAs that were available in the literature. We required that each system has both an Fe measurement ([Fe/H] $\lesssim$ $-2$) and a measurement of at least one of C, N, O, Al, Si, and S\@. Our sample spans a wide range of redshifts ($z_{abs} = 1.8-5.9$), neutral hydrogen column densities ($\log{N_{HI}} = 20.1-21.9$ in cm$^{-2}$), and metallicities ($-3.5 < {\rm [Fe/H]} < -1.9$). We re-normalized some of the solar abundances the datasets used to the \citet{asplund+2009} solar scale from older scales (e.g., \citealt{lodders_2003} or \citealt{asplund_2005}).

The majority ($\sim60\%$) of sources in our sample have high-resolution spectroscopic measurements. The observations were mainly split between the High Resolution Echelle Spectrometer (HIRES) \citep{vogt+1994} on Keck I and the Ultraviolet and Visual Echelle Spectrograph (UVES) \citep{dekker+2000} on the Very Large Telescope (VLT) UT2 at the European Southern Observatory.  HIRES typically covers a spectral range of 4000--8000~\AA\ with a resolution $R > 30,000$. UVES has a spectral range of 3000--8000~\AA\ between its red and blue configuration with a spectral resolution of R $\ge 40,000$.

The other $\sim40\%$ of our sample have medium-resolution spectroscopic measurements. Their spectra were obtained using the Echelle Spectrometer and Imager (ESI) \citep{sheinis+2002} on Keck II\@ at a resolution of $R \approx 5000$ spanning a spectral range of 3900--10900~\AA\@.

Additionally, two sources in our sample had spectra  obtained by the MIKE \citep{bernstein+2003} echelle spectrograph on the 6.5 m Magellan Clay telescope at Las Campanas with $R = 22,000-28,000$ covering a spectral range 3221--7420~\AA, and the XSHOOTER \citet{vernet+2011} spectrograph on VLT UT2 with $R \sim 8500$ covering a spectral range 3000 \AA\ -- 2.5$\mu$m.

The majority of our sample has appeared in multiple surveys. The first DLA survey was conducted in 1986 by \citet{wolfe+1986}, who compiled QSO candidates from the literature and followed them up at Lick Observatory. These sources, among others, were subsequently followed up by several authors once HIRES was commissioned in 1994 \citep[e.g.,][]{lu+1998, prochaska+2001, prochaska+2002a}. More sources were followed up, and discovered, following the commissioning of UVES in 2000 \citep[e.g.,][]{molaro+2000, ellison+2001, molaro+2001, dessauges-zavadsky+2001, levshakov+2002, lopez+2002, pettini+2002, centurion+2003, dessauges-zavadsky+2003}. In 2000 the Sloan Digital Sky Survey (SDSS) began operation and its first QSO sample target list was released by \citet{richards+2002}. These sources were later followed up by SDSS low resolution spectroscopy allowing an easy and automated way to search for VMP DLA candidates. One method, adopted by \citet{cooke+2011_sources}, searched for candidates by requiring that only 3 metal lines be measurable for a system in their SDSS spectra. The other surveys included in our sample are the UCSD HIRES DLA Survey \citep{prochaska_2007}, the Keck ESI MP DLA Survey \citep{penprase+2010}, and the \edit1{ESO UVES Advanced Data Products Quasar Sample \citep{zafar+2014}}. The rest of the sources are compiled from the following authors and references therein: \citet{pettini+2008, petitjean+2008, ellison+2010, srianand+2010, cooke+2011_sources, cooke+2011_carbon, cooke+2012, cooke+2013, cooke+2014, cooke+2015, berg+2016, cooke+2017, d'odorico+2018, welsh+2019, welsh+2020}.

Many ($\sim40\%$) systems in our compilation were observed by different authors resulting in multiple abundance measurements for individual systems. We defaulted to measurements from high resolution spectra, then those with the smallest reported uncertainties. In cases when the uncertainties were comparable we chose abundances derived/compiled in \citet[][]{cooke+2011_carbon,cooke+2011_sources,cooke+2012,cooke+2013,cooke+2014,cooke+2015,cooke+2016,cooke+2017,welsh+2019,welsh+2020}. The abundances used in our analysis are summarized in \textbf{\textnormal{Table \ref{tab:main_table} and will be available as an electronic table.%\edit2{The sample used for our analysis, in which we select a preferred measurement for each system (shown in Table \ref{tab:main_table}, will be available as an electronic table.
\footnote{The full compilation, which includes all abundance measurements for a system (including the measurements that were not used in our analysis), is available at  \dataset[https://github.com/evanhazey/CCSNe-Constraints-via-VMP-DLAs]{https://github.com/evanhazey/CCSNe-Constraints-via-VMP-DLAs}.}}}
%CCSNe-VMP-DLAs

%% LaTeX deluxetable generator for the AASTeX package.
%% Written by Greg Schwarz (5/1/2001).

%% Table generated: Thu Jun 10 12:58:21 2021
%% The values (usually only l,r and c) in the last part of
%% \begin{deluxetable}{} command tell LaTeX how many columns
%% there are and how to align them.
\startlongtable
\begin{longrotatetable}
\begin{deluxetable*}{lccccccccccccc}
\label{tab:main_table}
\movetabledown=10mm
\centerwidetable
%\tablewidth{2pt}
\tabletypesize{\scriptsize}
%% Keep a portrait orientation

%% Over-ride the default font size
%% Use Default (12pt)

%% Use \tablewidth{?pt} to over-ride the default table width.
%% If you are unhappy with the default look at the end of the
%% *.log file to see what the default was set at before adjusting
%% this value.

%% This is the title of the table.
\tablecaption{Metal Summary of VMP DLAs}

%% This command over-rides LaTeX's natural table count
%% and replaces it with this number.  LaTeX will increment 
%% all other tables after this table based on this number
%\tablenum{1}

%% The \tablehead gives provides the column headers.  It
%% is currently set up so that the column labels are on the
%% top line and the units surrounded by ()s are in the 
%% bottom line.  You may add more header information by writing
%% another line between these lines. For each column that requries
%% extra information be sure to include a \colhead{text} command
%% and remember to end any extra lines with \\ and include the 
%% correct number of &s.
\tablehead{\colhead{QSO} & \colhead{$z_{abs}$} & \colhead{$\log{N_{\rm HI}}$} & \colhead{[Fe/H]} & \colhead{[C/H]} & \colhead{[N/H]} & \colhead{[O/H]} & \colhead{[Al/H]} & \colhead{[Si/H]} & \colhead{[S/H]} & \colhead{Instrument} & \colhead{Ref} \\ 
\colhead{} & \colhead{} & \colhead{($cm^{-2}$)} & \colhead{} & \colhead{} & \colhead{} & \colhead{} & \colhead{} & \colhead{} & \colhead{} & \colhead{} & \colhead{} & \colhead{} & \colhead{} }

%% All data must appear between the \startdata and \enddata commands
\startdata
B0027-1836&2.402&21.75$\pm$0.1&-2.28$\pm$0.02$^d$&\nodata&\nodata&\nodata&\nodata&-1.59$\pm$0.03$^e$&-1.64$\pm$0.1&UVES&16\\ 
B1232+0815&2.3377&20.9$\pm$0.08&-1.96$\pm$0.08$^d$&\nodata&\nodata&\nodata&\nodata&-1.35$\pm$0.05$^e$&-1.21$\pm$0.12&UVES&18,24\\ 
BR0951-04&4.2029&20.4$\pm$0.1&$<$-2.6&\nodata&\nodata&\nodata&\nodata&-2.59$\pm$0.03&\nodata&HIRES&3,8\\ 
BR1202-07&4.3829&20.6$\pm$0.14&-2.22$\pm$0.12&\nodata&\nodata&\nodata&\nodata&-1.78$\pm$0.02&\nodata&HIRES&1\\ 
BR2237-0607&4.0803&20.52$\pm$0.11&-2.17$\pm$0.12&\nodata&\nodata&\nodata&\nodata&-1.84$\pm$0.02&\nodata&HIRES&1\\ 
BRI1108-07&3.6076&20.5$\pm$0.1&-2.15$\pm$0.01&\nodata&\nodata&\nodata&\nodata&-1.77$\pm$0.001&\nodata&HIRES&5,8\\ 
BRI1346-03&3.736&20.72$\pm$0.1&$<$-1.91&\nodata&\nodata&\nodata&\nodata&-2.28$\pm$0.01$^e$&\nodata&HIRES&8\\ 
BRJ0426-2202&2.9831&21.5$\pm$0.15&-2.78$\pm$0.06&\nodata&\nodata&\nodata&\nodata&$<$-2.0&\nodata&ESI&13\\ 
CTQ247&2.6215&20.47$\pm$0.1&-2.4$\pm$0.02&\nodata&\nodata&\nodata&\nodata&-2.01$\pm$0.06&\nodata&ESI&13\\ 
HS0741+4741&3.017&20.48$\pm$0.1&-1.93$\pm$0.01$^d$&\nodata&\nodata&\nodata&\nodata&-1.64$\pm$0.01$^e$&-1.6$\pm$0.1&ESI,HIRES&10,8\\ 
HS1132+2243&2.7835&21.0$\pm$0.07&-2.5$\pm$0.01&\nodata&\nodata&\nodata&\nodata&-2.05$\pm$0.14&\nodata&ESI&13\\ 
J0035-0918 &2.3401&20.55$\pm$0.1&-3.04$\pm$0.12&-1.51$\pm$0.18&-2.87$\pm$0.12&-2.28$\pm$0.13&-3.26$\pm$0.11&-2.65$\pm$0.11&\nodata&HIRES,UVES&25, 31,29\\ 
J0140-0839 &3.6966&20.75$\pm$0.15&-3.45$\pm$0.24&-3.05$\pm$0.17&$<$-4.20&-2.75$\pm$0.15&-3.37$\pm$0.16&-2.75$\pm$0.17&$<$-2.54&HIRES,UVES&21,25\\ 
J0255+00&3.9146&21.3$\pm$0.05&-2.08$\pm$0.09&\nodata&\nodata&\nodata&\nodata&\nodata&-1.71$\pm$0.01&HIRES&8\\ 
J0307-4945 &4.46658&20.67$\pm$0.09&-1.93$\pm$0.19&\nodata&-2.93$\pm$0.15&-1.45$\pm$0.19&-1.75$\pm$0.11&-1.5$\pm$0.11&\nodata&UVES&6\\ 
J0311-1722 &3.734&20.3$\pm$0.06&$<$-2.01&-2.71$\pm$0.1&$<$-3.06&-2.29$\pm$0.1&\nodata&-2.5$\pm$0.09&\nodata&UVES&26\\ 
J0831+3358 &2.30364&20.25$\pm$0.15&-2.39$\pm$0.16&\nodata&$<$-3.30&-2.01$\pm$0.16&-2.5$\pm$0.16&-2.01$\pm$0.16&\nodata&HIRES&26,22\\ 
J0903+2628 &3.0776&20.32$\pm$0.05&$<$-2.81&-3.43$\pm$0.03$^h$&\nodata&-3.05$\pm$0.05&\nodata&-3.21$\pm$0.02$^i$&\nodata&HIRES&34\\ 
J0953-0504&4.20287&20.55$\pm$0.1&-2.98$\pm$0.21&-3.05$\pm$0.1&$<$-2.84&-2.55$\pm$0.1&\nodata&-2.7$\pm$0.1&$<$-1.78&HIRES,UVES&29\\ 
J1001+0343 &3.07841&20.21$\pm$0.05&-3.18$\pm$0.15&-3.06$\pm$0.05&$<$-3.54&-2.65$\pm$0.05&\nodata&-2.86$\pm$0.05&\nodata&HIRES,UVES&26,31\\ 
J1037+0139 &2.70487&20.5$\pm$0.08&-2.44$\pm$0.08&\nodata&-3.06$\pm$0.09&-2.13$\pm$0.09&-2.62$\pm$0.09&-2.04$\pm$0.09&\nodata&UVES&26\\ 
J1111+1332&2.27094&20.39$\pm$0.04&-2.27$\pm$0.04&-2.1$\pm$0.11$^g$&\nodata&-1.92$\pm$0.08&\nodata&-1.95$\pm$0.02$^f$&\nodata&HIRES,UVES&31\\ 
J1113-1533&3.2665&21.23$\pm$0.05&-2.08$\pm$0.03$^d$&\nodata&\nodata&\nodata&\nodata&\nodata&-1.73$\pm$0.07&HIRES,UVES&30\\ 
J1337+3152&3.1735&21.3$\pm$0.09&-1.93$\pm$0.05$^d$&\nodata&\nodata&\nodata&\nodata&-1.37$\pm$0.08$^e$&-1.34$\pm$0.17&UVES&23\\ 
J1337+3153 &3.16768&20.41$\pm$0.15&-2.74$\pm$0.3&-2.86$\pm$0.16&$<$-3.44&-2.67$\pm$0.17&-2.85$\pm$0.16&-2.68$\pm$0.16&\nodata&UVES&23\\ 
J1340+1106 &2.50792&20.09$\pm$0.05&-2.07$\pm$0.05&\nodata&-3.12$\pm$0.06&-1.76$\pm$0.06&-2.26$\pm$0.05&-1.85$\pm$0.05&-1.81$\pm$0.05&HIRES,UVES&25,26\\ 
J1358+0349&2.853054&20.27$\pm$0.02&$<$-3.25&\nodata&-3.58$\pm$0.11&-2.804$\pm$0.015&$<$-2.95&$<$-2.764&$<$-2.64&HIRES&32\\ 
J1358+6522&3.067295&20.47$\pm$0.07&-2.84$\pm$0.03&-2.25$\pm$0.1&-3.68$\pm$0.14&-2.22$\pm$0.05&-2.99$\pm$0.03&-2.58$\pm$0.03&-2.5$\pm$0.09&HIRES&27\\ 
J1419+0829 &3.04973&20.4$\pm$0.03&-2.33$\pm$0.04&\nodata&-2.95$\pm$0.04&-1.92$\pm$0.04&\nodata&-2.08$\pm$0.03&\nodata&UVES&26\\ 
J1558-0031 &2.70262&20.67$\pm$0.05&-2.03$\pm$0.05&\nodata&-2.04$\pm$0.05&-1.5$\pm$0.05&\nodata&-1.94$\pm$0.05&\nodata&HIRES,MIKE&15\\ 
J1558+4053 &2.55332&20.3$\pm$0.04&-2.7$\pm$0.07&-2.51$\pm$0.07&-3.47$\pm$0.08&-2.45$\pm$0.06&-2.82$\pm$0.07&-2.49$\pm$0.04&\nodata&UVES&20\\ 
J2310+1855$^j$&5.938646&21.05$\pm$0.1&-3.08$\pm$0.12&\nodata&\nodata&\nodata&\nodata&-2.86$\pm$0.14&\nodata&XSHOOTER&35\\ 
J2321+1421&2.5731&20.7$\pm$0.05&-2.02$\pm$0.03$^d$&\nodata&\nodata&\nodata&\nodata&-1.76$\pm$0.04$^e$&$<$-2.22&UVES&21\\ 
PC0953+47&4.2442&20.9$\pm$0.15&-2.55$\pm$0.08&\nodata&\nodata&\nodata&\nodata&-2.16$\pm$0.03&\nodata&ESI&13\\ 
PKS1354-17&2.7799&20.3$\pm$0.15&-2.36$\pm$0.08&\nodata&\nodata&\nodata&\nodata&-1.83$\pm$0.05&\nodata&HIRES&13\\ 
PSS0808+52&3.1132&20.65$\pm$0.07&-2.01$\pm$0.04&\nodata&\nodata&\nodata&\nodata&-1.51$\pm$0.11&\nodata&ESI&9,13\\ 
PSS0957+33&4.1798&20.7$\pm$0.1&-2.1$\pm$0.05&\nodata&\nodata&\nodata&\nodata&-1.67$\pm$0.01&\nodata&ESI,HIRES&8,9\\ 
PSS1248+31&3.697&20.63$\pm$0.07&-2.23$\pm$0.05&\nodata&\nodata&\nodata&\nodata&-1.69$\pm$0.01&\nodata&ESI&9,13\\ 
PSS1506+5220&3.2244&20.67$\pm$0.07&-2.48$\pm$0.04&\nodata&\nodata&\nodata&\nodata&-2.3$\pm$0.02&\nodata&ESI&13\\ 
PSS1715+3809&3.3407&21.05$\pm$0.13&-2.82$\pm$0.04&\nodata&\nodata&\nodata&\nodata&$<$-2.08&\nodata&ESI&14\\ 
PSS1802+5616&3.8109&20.35$\pm$0.2&-2.2$\pm$0.11&\nodata&\nodata&\nodata&\nodata&-2$\pm$0.1&\nodata&ESI&14\\ 
PSS2323+2758&3.6845&20.95$\pm$0.1&-3.08$\pm$0.12&\nodata&\nodata&\nodata&\nodata&-2.56$\pm$0.03&\nodata&ESI&13\\ 
Q0000-2620&3.3901&21.41$\pm$0.08&-2.04$\pm$0.03$^d$&\nodata&-2.54$\pm$0.08&-1.68$\pm$0.13&\nodata&-1.86$\pm$0.02$^e$&-1.83$\pm$0.09&HIRES,UVES&7,4\\ 
Q0112-306 &2.41844&20.5$\pm$0.08&-2.64$\pm$0.09&\nodata&-3.17$\pm$0.09&-2.24$\pm$0.11&\nodata&-2.39$\pm$0.08&\nodata&UVES&19\\ 
Q0913+072 &2.61843&20.34$\pm$0.04&-2.82$\pm$0.04&-2.79$\pm$0.06&-3.88$\pm$0.13&-2.4$\pm$0.04&-3$\pm$0.05&-2.55$\pm$0.04&\nodata&HIRES,UVES&28,20\\ 
Q0930+2858&3.235&20.3$\pm$0.1&-2.1$\pm$0.02$^d$&\nodata&\nodata&\nodata&\nodata&-1.92$\pm$0.02$^e$&\nodata&HIRES&10\\ 
Q1021+30&2.9489&20.7$\pm$0.1&-2.19$\pm$0.01&\nodata&\nodata&\nodata&\nodata&-1.91$\pm$0.02&\nodata&HIRES&8,13\\ 
Q1108-077 &3.60767&20.37$\pm$0.07&-1.96$\pm$0.07&\nodata&$<$-3.36&-1.69$\pm$0.08&\nodata&-1.54$\pm$0.07&\nodata&UVES&19\\ 
Q1331+17&1.7764&21.14$\pm$0.08&-2.05$\pm$0.001&\nodata&\nodata&\nodata&\nodata&-1.39$\pm$0.001&\nodata&HIRES&3,8\\ 
Q1337+11&2.7959&20.95$\pm$0.1&-2.41$\pm$0.02&\nodata&\nodata&\nodata&\nodata&-1.69$\pm$0.07&\nodata&ESI,HIRES&13,17\\ 
Q1409+095&2.4562&20.54$\pm$0.1&-2.33$\pm$0.02&\nodata&\nodata&\nodata&\nodata&-1.99$\pm$0.02&\nodata&UVES&12\\ 
Q1451+123&2.469&20.39$\pm$0.1&-2.49$\pm$0.05&\nodata&\nodata&\nodata&\nodata&-2.1$\pm$0.1&\nodata&UVES&12\\ 
Q1946+7658 &2.8443&20.27$\pm$0.06&-2.5$\pm$0.06&\nodata&-3.51$\pm$0.07&-2.14$\pm$0.06&\nodata&-2.18$\pm$0.06&\nodata&ESI&10\\ 
Q2059-360 &3.08293&20.98$\pm$0.08&-1.97$\pm$0.08&\nodata&-2.86$\pm$0.08&-1.58$\pm$0.09&\nodata&-1.63$\pm$0.09&\nodata&UVES&19\\ 
Q2206-199 &2.07624&20.43$\pm$0.04&-2.57$\pm$0.04&-2.45$\pm$0.05&-3.47$\pm$0.06&-2.07$\pm$0.05&-2.69$\pm$0.04&-2.29$\pm$0.04&\nodata&UVES&20\\ 
Q2223+20&3.1192&20.3$\pm$0.1&-2.4$\pm$0.04&\nodata&\nodata&\nodata&\nodata&-2.2$\pm$0.04&\nodata&ESI&13\\ 
Q2348-01&2.6147&21.3$\pm$0.1&-2.26$\pm$0.09&\nodata&\nodata&\nodata&\nodata&-1.95$\pm$0.07&\nodata&HIRES&8\\ 
Q2348-14&2.2794&20.56$\pm$0.08&-2.27$\pm$0.02&\nodata&\nodata&\nodata&\nodata&-1.89$\pm$0.02&\nodata&HIRES&3,8\\ 
S0759+3129$^j$&3.0346&20.6$\pm$0.1&-2.3$\pm$0.15$^a$&\nodata&\nodata&\nodata&-2.621$\pm$0.2$^c$&-2$\pm$0.2$^c$&\nodata&ESI&22\\ 
S0928+0939$^j$&2.9098&20.75$\pm$0.15&-2.16$\pm$0.2$^b$&-2.449$\pm$0.15$^b$&\nodata&\nodata&\nodata&\nodata&\nodata&ESI&22\\ 
S0955+4116$^j$&3.2801&20.1$\pm$0.1&$<$-2.33&-2.859$\pm$0.15$^b$&\nodata&-2.82$\pm$0.15$^b$&$<$-2.731&-2.72$\pm$0.15$^b$&\nodata&ESI&22\\ 
S1001+0343$^j$&3.0785&20.15$\pm$0.1&$<$-2.35&-2.889$\pm$0.15$^b$&\nodata&-2.93$\pm$0.15$^b$&-2.811$\pm$0.15$^b$&-2.91$\pm$0.15$^b$&\nodata&ESI&22\\ 
S1003+5520$^j$&2.5024&20.35$\pm$0.15&-2.9$\pm$0.2$^c$&$<$-2.269&\nodata&$<$-2.21&-2.751$\pm$0.15$^b$&-2.1$\pm$0.2$^c$&\nodata&ESI&22\\ 
S1031+4055$^j$&2.5686&20.55$\pm$0.1&-2.21$\pm$0.15$^b$&$<$-2.619&\nodata&$<$-2.21&\nodata&$<$-1.63&\nodata&ESI&22\\ 
S1048+3911$^j$&2.2957&20.7$\pm$0.1&-2.49$\pm$0.15$^b$&-2.829$\pm$0.1$^a$&\nodata&\nodata&-2.481$\pm$0.2$^c$&-2.28$\pm$0.2$^b$&\nodata&ESI&22\\ 
S1108+1209$^j$&3.3964&20.55$\pm$0.15&$<$-2.32&\nodata&\nodata&$<$-2.61&\nodata&\nodata&\nodata&ESI&22\\ 
S1219+1603$^j$&3.0037&20.35$\pm$0.1&-2.12$\pm$0.15$^b$&\nodata&\nodata&-2.59$\pm$0.2$^c$&\nodata&-2.05$\pm$0.15$^b$&\nodata&ESI&22\\ 
S1251+4120$^j$&2.7296&21.1$\pm$0.1&-2.38$\pm$0.2$^c$&\nodata&\nodata&\nodata&-2.851$\pm$0.2$^c$&-2.7$\pm$0.2$^c$&\nodata&ESI&22\\ 
S1305+2902$^j$&2.3865&20.25$\pm$0.1&-2.82$\pm$0.15$^b$&-2.499$\pm$0.2$^c$&\nodata&-2.9$\pm$0.15$^b$&-2.821$\pm$0.15$^b$&-2.51$\pm$0.15$^b$&\nodata&ESI&22\\ 
S1325+1255$^j$&3.5507&20.5$\pm$0.15&$<$-2.30&-2.579$\pm$0.2$^c$&\nodata&\nodata&$<$-2.051&-2.49$\pm$0.15$^b$&\nodata&ESI&22\\ 
S1350+5952$^j$&2.7558&20.65$\pm$0.1&-2.62$\pm$0.15$^b$&\nodata&\nodata&\nodata&-2.521$\pm$0.2$^c$&\nodata&\nodata&ESI&22\\ 
S1440+0637$^j$&2.8246&20.2$\pm$0.1&-2.24$\pm$0.15$^b$&$<$-2.469&\nodata&$<$-1.96&-2.421$\pm$0.15$^b$&-2.11$\pm$0.11$^a$&\nodata&ESI&22\\ 
S1456+0407$^j$&2.6736&20.35$\pm$0.1&-2.92$\pm$0.15$^b$&\nodata&\nodata&-2.56$\pm$0.2$^c$&-1.951$\pm$0.2$^c$&-2.44$\pm$0.15$^b$&\nodata&ESI&22\\ 
S1557+2320$^j$&3.5383&20.65$\pm$0.1&-2.64$\pm$0.2$^c$&$<$-2.849&\nodata&-2.21$\pm$0.11$^a$&-2.671$\pm$0.2$^c$&-2.11$\pm$0.2$^c$&\nodata&ESI&22\\ 
S1637+2901$^j$&3.4956&20.7$\pm$0.1&-2.43$\pm$0.15$^b$&\nodata&\nodata&-3.17$\pm$0.2$^c$&-2.941$\pm$0.15$^b$&-2.87$\pm$0.15$^b$&\nodata&ESI&22\\ 
S1654+3509$^j$&2.8113&20.1$\pm$0.1&-2.04$\pm$0.11$^a$&$<$-1.739&\nodata&\nodata&-1.521$\pm$0.11$^a$&-1.71$\pm$0.2$^c$&\nodata&ESI&22\\ 
S1709+3417$^j$&3.0104&20.4$\pm$0.1&-2.02$\pm$0.15$^b$&\nodata&\nodata&\nodata&-2.211$\pm$0.2$^c$&\nodata&\nodata&ESI&22\\ 
S1717+5802$^j$&3.0461&20.25$\pm$0.1&-2.4$\pm$0.11$^a$&\nodata&\nodata&\nodata&\nodata&-2.02$\pm$0.15$^b$&\nodata&ESI&22\\ 
S2114-0632$^j$&4.1262&20.4$\pm$0.15&$<$-2.46&-2.669$\pm$0.2$^c$&\nodata&-2.44$\pm$0.2$^c$&$<$-3.161&-2.75$\pm$0.15$^b$&\nodata&ESI&22\\ 
\enddata

    \tablerefs{1: \citet[][]{lu+1996}; 2: \citet[][]{prochaska+1997a}; 3: \citet[][]{prochaska+1999}; 4: \citet[][]{molaro+2000}; 5: \citet[][]{prochaska+2000}; 6: \citet[][]{dessauges-zavadsky+2001}; 7: \citet[][]{molaro+2001}; 8: \citet[][]{prochaska+2001}; 9: \citet[][]{prochaska+2001a}; 10: \citet[][]{prochaska+2002a}; 11: \citet[][]{dessauges-zavadsky+2001}; 12: \citet[][]{ledoux+2003}; 13: \citet[][]{prochaska+2003c}; 14: \citet[][]{prochaska+2003d}; 15: \citet[][]{omeara+2006}; 16: \citet[][]{noterdaeme+2007}; 17: \citet[][]{prochaska_2007}; 18: \citet[][]{noterdaeme+2008}; 19: \citet[][]{petitjean+2008}; 20: \citet[][]{pettini+2008}; 21: \citet[][]{ellison+2010}; 22: \citet[][hereafter P10]{penprase+2010}; 23: \citet[][]{srianand+2010}; 24: \citet[][]{balashev+2011}; 25: \citet[][]{cooke+2011_sources}; 26: \citet[][]{cooke+2011_carbon}; 27: \citet[][]{cooke+2012}; 28: \citet[][]{cooke+2014}; 29: \citet[][]{dutta+2014}; 30: \citet[][hereafter Z14]{zafar+2014}; 31: \citet[][hereafter C15]{cooke+2015}; 32: \citet[][]{cooke+2016}; 33: \citet[][]{morrison+2016}; 34: \citet[][hereafter C17]{cooke+2017}; 35: \citet[][hereafter D18]{d'odorico+2018}; 36: \citet[][hereafter W19]{welsh+2019}; 37: \citet[][]{welsh+2020}}
    \tablenotetext{a}{Uncertainty less than 0.11 dex (P10)}
    \tablenotetext{b}{Uncertainty between 0.11 and 20 dex (P10)}
    \tablenotetext{c}{Uncertainty greater than 0.20 dex (P10)}
    \tablenotetext{d}{Reporting uncertainty on $N(Fe)$ (Z14)}
    \tablenotetext{e}{Reporting uncertainty on $N(Si)$ (Z14)}
    \tablenotetext{f}{Reporting uncertainty on [Si/Fe] (C15)}
    \tablenotetext{g}{Reporting uncertainty on [C/O] (C17)}
    \tablenotetext{h}{Reporting uncertainty on [C/O] (W19)}
    \tablenotetext{i}{Reporting uncertainty on [Si/O] (W19)}
    \tablenotetext{j}{Name shortened from original (P10, D18)}
    \tablecomments{This table is published in the machine-readable format}
    %\label{tab:main_table}
\end{deluxetable*}
\end{longrotatetable}

\section{Empirical Constraints on CCSN Yields} \label{sec:empricial_yields}
%\subsection{Abundance Ratios in VMP DLAs} \label{sec:abundance_ratios}
We place empirical constraints on the IMF-averaged yields of zero to low-metallicity core-collapse supernovae by analysis of the observed abundances of VMP DLAs.  Specifically, we find the values of the low-metallicity plateaus in abundance ratios for the most readily measurable elements (C, N, O, Al, Si, S, and Fe). The first enriching events of system are CCSNe, followed by delayed enriching events, such as winds from AGB stars and Type Ia SN (SNe Ia). These processes can be disentangled in the space of [X/Fe] vs.\ [Fe/H] because [Fe/H] can be assumed to monotonically increase with time (assuming a relatively smooth star formation history), especially at low metallicity. At the earliest times, or lowest [Fe/H], the ratios should reflect the yields from CCSNe alone.  The abundance ratios appear constant (a plateau in [X/Fe] vs.\ [Fe/H]) at low metallicities because there is only one type of enrichment (CCSNe, though not necessarily a single CCSN). At later times, or higher [Fe/H], the ratios reflect a combination of yields from CCSNe and delayed processes. The introduction of a new enrichment source leads to a change in [X/Fe] (a ``knee'' in [X/Fe] vs.\ [Fe/H]) at the value of [Fe/H] when the new sources turn on. In the case of the [$\alpha$/Fe] ratio, CCSNe produce both $\alpha$-elements and Fe-peak elements %(massive contribute both roughly the same proportion ) 
whereas SNe Ia produce mainly Fe-peak elements and little to no $\alpha$-elements. Therefore, locating the [X/Fe] plateau gives the abundance ratio of CCSNe.

\begin{figure*}
    \centering
    \includegraphics[scale=0.69]{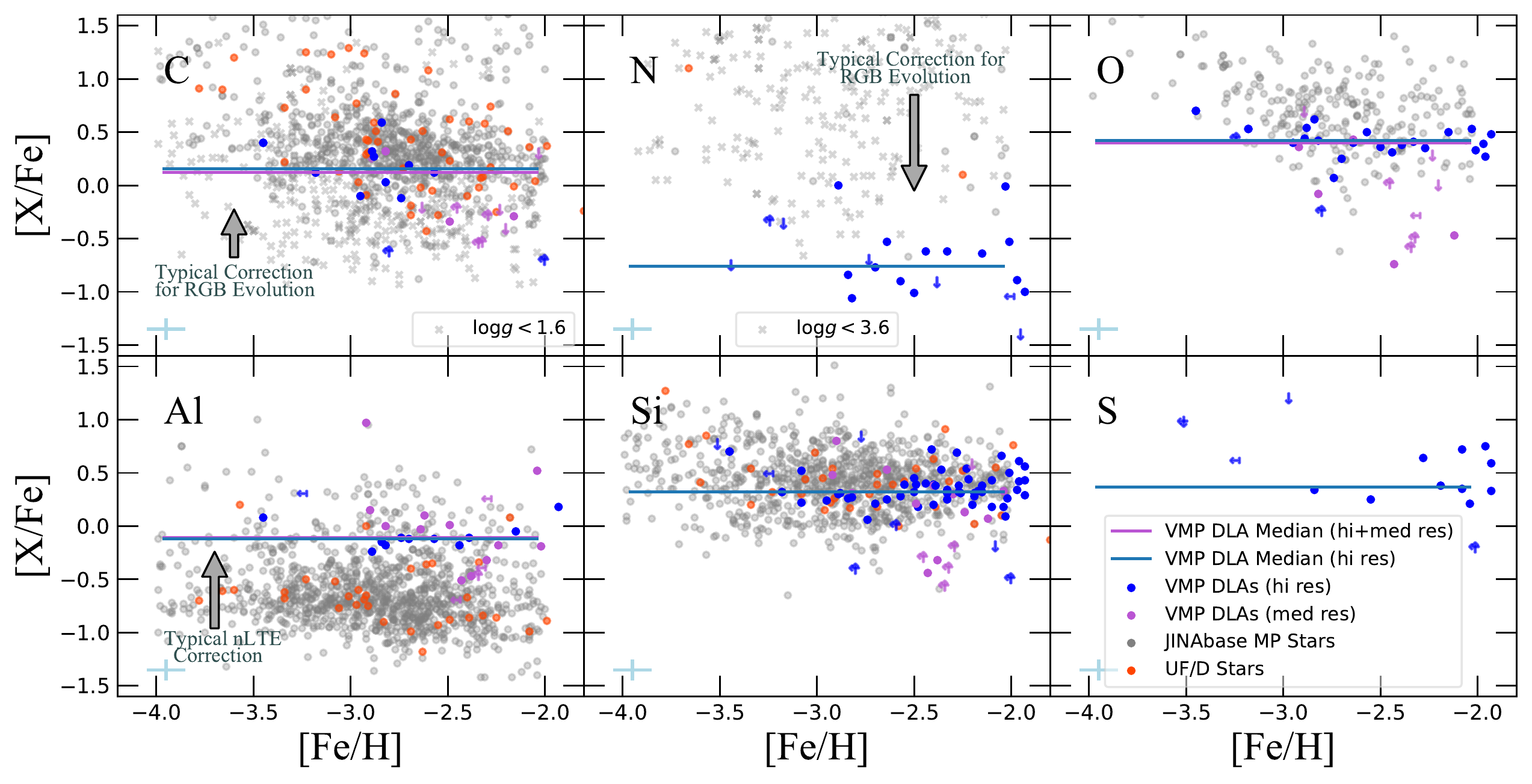}
    \caption{C, N, O, Al, Si, and S abundance ratios as a function of [Fe/H]. The blue (purple) points are VMP DLAs whose abundances were measured from high (medium) resolution spectra.  The blue horizontal bars are the medians of the VMP DLA abundance ratios from high resolution sources, and the purple horizontal bars are the medians of the VMP DLA abundance ratios from high+medium resolution sources.  Typical VMP DLA uncertainties (0.1 dex) are shown in the bottom left corner of each subplot. The grey smaller points are metal-poor stars from the JINA database \citep{abohalima+2018}; the small gray x's are giants whose surface abundances have been altered by RGB evolution. The red smaller points are from the ultra-faint/dwarf galaxy star compilation from Alexander Ji (See Section \ref{sec:dla_stars_comparison}). The gray arrows in the subplots show the typical corrections \citep{placco+2014,nordlander+2017} for the stated physical processes, to scale (See Section \ref{sec:dla_stars_comparison} for discussion).  The corrections vary widely from star to star. UFD Compilation References: \citet{koch+2008,feltzing+2009,frebel+2010,simon+2010,norris+2010b,lai+2011,gilmore+2018,koch+2013,frebel+2014,ishigaki+2014,roederer+2014,francois+2016,ji+2016a,ji+2016b,roederer+2016,hansen+2017,kirby+2017,nagasawa+2018,chiti+2018,spite+2018,marshall+2019,ji+2019}.}
    \label{fig:XFe_FeH}
\end{figure*}

DLAs have been shown to exhibit a [$\alpha$/Fe] knee at [Fe/H] $\approx$ $-2$ \citep{cooke+2015} preceded by a plateau, implying that the VMP (${\rm [Fe/H]} \leq -2$) DLAs in our sample are in this plateau. To identify [X/Fe] values of these plateaus, and hence to obtain the abundance ratio of CCSN yields, we take the {\it median} of the abundance ratios observed in VMP DLAs, shown in Figure \ref{fig:XFe_FeH}. We chose to use medians rather than means for several reasons.  First, the mean is different depending on whether it is taken in logarithmic space (e.g., bracket notation like [O/Fe]) or linear space (e.g., mass ratios like $M({\rm O})/M({\rm Fe})$).  The median is the same in either space.  More importantly, the median is less sensitive to outliers, which might result from systematic uncertainties that are difficult to correct for (e.g., spurious instrumental errors, inconsistencies in abundance determinations, ionization corrections). 

\begin{figure*}
    \centering
    \includegraphics[scale=0.69]{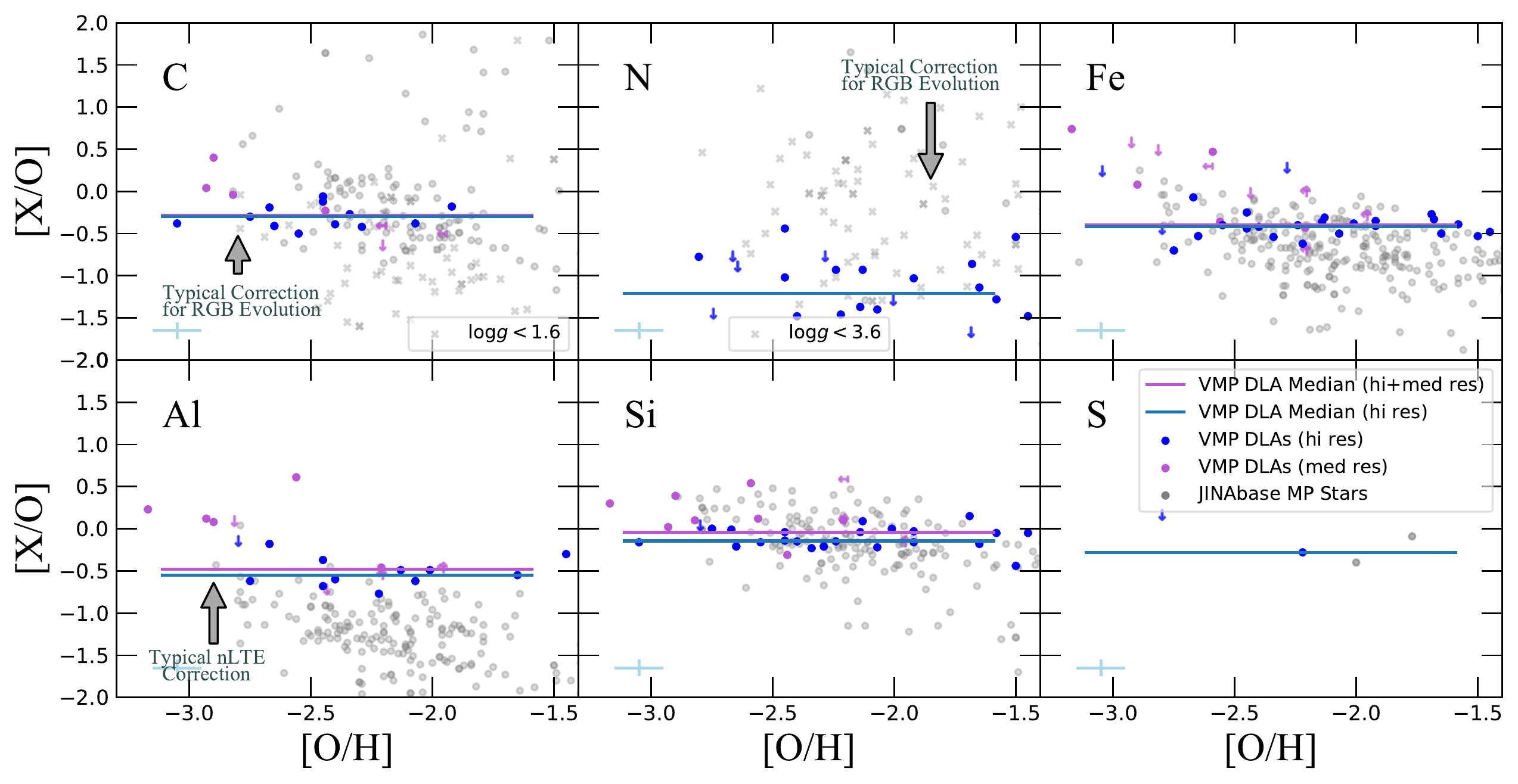}
    \caption{\edit1{C, N, Fe, Al, Si, and S abundance ratios as a function of [O/H]. Same symbols as Figure \ref{fig:XFe_FeH}}}
    \label{fig:XO_OH}
\end{figure*}

The median calculations begin by first finding the median for sources that are doubly bounded (i.e., no upper or lower limits). Then, using this preliminary median, we find all upper limits that are below it and all lower limits that are above it. Finally, we recalculate the median, and associated 68\% confidence intervals (in log space), using the doubly bounded sources and the aforementioned meaningful upper/lower limits. The medians are shown as the colored horizontal lines in Figures \ref{fig:XFe_FeH} and \ref{fig:XO_OH}.% and tabulated in Table \ref{tab:med_abundance_ratios} with 1-$\sigma$ uncertainties. 
The figures also contain the abundance ratios of metal-poor stars, which we discuss in Section \ref{sec:dla_stars_comparison}. %We average the abundances in linear space. 
%An abundance in bracket notation, like [X/Fe], is a logarithmic quantity. However, theoretical supernova yields and chemical evolution models conventionally use linear values.  Therefore, I average the abundance ratios in linear space: $\overline{\rm [X/Fe]} = \log \left( {\rm mean}(10^{\rm [X/Fe]}) \right)$. I list the averages in Tabl   e \ref{tab:avg_ratios}.
%The median of the abundance ratios with respect to Fe are shown in Table \ref{tab:med_XFe}. 

\edit1{Abundances measured from medium resolution spectra}\textbf{\textnormal{, $R < 10,000$,}} \edit1{can contain uncertainties not present in high resolution spectra} \textbf{\textnormal{such as line blending and/or hidden saturation}}\edit1{, which can result in their abundances being inaccurate. To account for this, we calculate the median of each abundance ratio twice.} First, we use all sources in the sample (i.e., with abundances measured from both high resolution spectra and medium resolution spectra; purple horizontal line(s) in Figures \ref{fig:XFe_FeH} and \ref{fig:XO_OH}). Second, we use abundances measured only from high resolution spectra (blue horizontal line(s) in Figures \ref{fig:XFe_FeH} and \ref{fig:XO_OH}). The differences between the medians were always $<$0.1 dex and as small as 0.01 dex. The 1-$\sigma$ uncertainties decreased by about 0.1 dex when culling the sample to high resolution sources only. The uncertainty for [Al/O] decreased by 0.2 dex.

The [O/Fe] and [Si/Fe] abundance ratios contain a subset of sources between [Fe/H]$\sim$-2.5-- -2 whose abundances are $>$0.5 dex below the bulk trends. The majority ($\sim$70\%) of the sources come from \citet[][hereafter P10]{penprase+2010}, were measured from medium resolution spectra, and have only upper limits on [Fe/H].%, span a wide range of \ion{H}{1} column density ($\log{N_{HI}} = 20.1-21.15$), and span a redshift range of $z_{abs}$ 3.0--4.1. 
There are five sources who have both low [O/Fe] and [Si/Fe] \edit1{(compared to the bulk trends)} and three of them come from P10. These were among the most oxygen poor sources ([O/H]$<$-2.6) and silicon poor sources ([Si/H]$<$-2.5) in the sample.

\edit1{Dust depletion has been shown to be minimal in DLAs at low-metallicity \citep[See Section \ref{sec:intro};][]{wolfe+2005,cooke+2011_sources}.} \textbf{\textnormal{While w\edit1{e cannot prove that there is no dust depletion in the VMP} DLAs, we \edit1{point to our work and that of others to argue that it is} negligible. Figure \ref{fig:XFe_FeH} shows the observed abundance ratios of \edit1{[S/Fe] and [Si/Fe]; S is an alpha element that is volatile (i.e., it does not easily condense onto dust grains similar to C, N, O, and Al, so its gas phase abundance is the same as its true abundance), and Si and Fe are both refractory elements (i.e., easily condense onto dust grains, Fe more so than Si ). %It also traces Fe which means that [S/Fe] should not trend with metallicity if dust is negligible;
%We observe no trend in [S/Fe] in Figure \ref{fig:XFe_FeH}. On the other hand,   As a result, 
If dust depletion is appreciable there should be a trend between [Si/Fe] and [S/Fe] with metallicity; we observe no trend}\textbf{\textnormal{ implying that the relative dust depletion between S, Si, and Fe is negligible}}\edit1{. Further, the stellar abundance ratio---which is not subject to dust corrections---at low metallicities in the Milky Way is [Si/Fe]~=~$\rm 0.37\pm 0.15$ \citep[][]{cayrel+2004}; we observe [Si/Fe]~=~$\rm 0.32 \pm 0.16$.}}} Finally, in recent work from \citet[][]{decia+2018}, the dust depletion of Si and Fe in DLAs was shown to be effectively zero at [Fe/H] $\sim$ -2 but as large as 1 dex at [Fe/H] $\sim$ 0. Other studies have shown the same trend of dust depletion decreasing as metallicities approach 1/100 solar \citep[][]{pettini+1997,vladilo+2002,akerman+2005,wolfe+2005,vladilo+2018}. \textbf{\textnormal{All suggest minimal effects of dust on the abundance ratios measured in the gas phase.}} %\edit2{Altogether, the lack of a trend in the observed abundance ratios [Si/Fe] and [S/Fe], and their similarity to stellar ratios shows that their relative dust depletion in VMP DLAs is negligible.}

%\edit1{That said, if dust were truly non-negligible then there is a systemic increase present in [C/Fe], [N/Fe], [O/Fe], [Al/Fe], [Si/Fe], and [S/Fe]. This is because if Fe is depleted, its gas phase abundance is an underestimate of its true abundance. Correcting for this would mean increasing the abundance of Fe because N(Fe)$_{total}$ = N(Fe)$_{gas}$ + N(Fe)$_{dust}$, which would decrease the aforementioned abundance ratios. This implies that if dust depletion is non-neglible, the current abundance ratios are systematically increased (because additional Fe is locked onto dust grains and removed from the gas phase abundance). We cannot quantify the magnitude of this increase so instead argue, using previously mentioned reasons, that the magnitude is smaller than the typical uncertainties of the VMP DLA abundance measurements ($<$0.1 dex)\citep[][]{vladilo+2002,wolfe+2005,decia+2018}. We discuss how our comparison to nucleosynthetic yields aids this argument in Section \ref{sec:theory_comparison}.}

In order to separate nucleosynthesis during the pre-SN evolution of the progenitor stars and during the explosion of the stars, we show the VMP DLA abundance ratios with respect to oxygen in Figure \ref{fig:XO_OH}. Oxygen provides insight into the pre-SN evolution of the star because it is synthesized primarily during hydrostatic burning.  On the other hand, Fe is synthesized during the SN explosion.  Intermediate elements, such as Si, are produced significantly in both hydrostatic and explosive nucleosynthesis. This distinction will play an important role in our comparison to theoretical yield models in Section \ref{sec:theory_comparison}. 

There should be a constant trend in the abundance ratios with respect to oxygen because oxygen is synthesized hydrostatically, the majority of the elements in Figure \ref{fig:XO_OH} are synthesized hydrostatically, and there is only one source of nucleosynthesis at the low metallicites that we are probing. In other words, most of the elements, except Fe, and perhaps Si and S, should be produced roughly in the same proportion in CCSNe. 

There is evidence suggesting that [C/O] (vs.\ [O/H]) decreases for VMP DLAs until a minimum is reached at [O/H]~$\sim$~-1.5 \citep[][]{pettini+2008,penprase+2010,cooke+2011_sources}. \edit1{This finding is based on extrapolating the behavior seen in red giants in the Milky Way halo whose [C/O] (vs.\ [O/H]) shows a decrease at low [O/H], a minimum at [O/H]~$\sim$~-1.5, and an increase to solar at high [O/H] \citep[e.g.,][]{akerman+2004,fabbian+2009}. This rise in [C/O] at low oxygen abundance has been interpreted as a PopIII signature owing to C enhancements from zero metallicity stars. But the C abundances for red giants can be uncertain due to the astration corrections necessary to infer their abundance \citep[e.g.,][]{smith+2006,placco+2014,Kirby+2015}. Also, C and O also could have 3D nLTE corrections as large as \textbf{\textnormal{-0.3 dex and -0.6 dex}} respectively. Recently, \citet[][]{amarsi+2019a, amarsi+2019b} calculated 3D nLTE corrections for C and O abundances in metal-poor stars and found that the downturn in [C/O] (at low oxygen abundance) is no longer present; an increase is seen instead. Interestingly, i}f one were to observe [C/O] (vs.\ [O/H]) from the VMP DLAs in isolation, a strong trend with [O/H] is not apparent; a weak trend may be present \citep[][]{cooke+2017,poudel+2019,berg+2021}. %Therefore, we argue there is a [C/O] plateau at these low metallicities in VMP DLAs.

Nitrogen is also affected by astration but it has been shown that [N/O] (vs.\ [O/H]) does not vary with oxygen abundances below [O/H]~$\sim -0.7$ for DLAs and instead reaches a plateau \citep[][]{petitjean+2008,pettini+2008,zafar+2014a}. This behavior is similar to what is found for [N/O] in local dwarf galaxies at the lowest [O/H] where for [O/H] $\lesssim -0.7$ there is a primary nitrogen plateau ([N/O]$\sim$~$-0.65$) then a rapid rise with increasing [O/H]; for DLAs the plateau at low [O/H] is more than 0.3 dex lower \citep[][]{Berg+2019}.

For these reasons, we use the same approach, rationale, and calculation to find the medians of these ratios with respect to oxygen, and interpret the medians as the yield ratios of CCSNe.

The medians of the abundance ratios with respect to oxygen in Figure \ref{fig:XO_OH} (and with respect to Fe), and their 1-$\sigma$ (68\% confidence interval) uncertainties, are listed in Table \ref{tab:med_abundance_ratios}. These medians reflect the median IMF-averaged CCSN yield for zero- and low-metallicity massive stars.

\begin{table}[htb]
\centering
\caption{Median Abundance Ratios of VMP DLAs} \label{tab:med_abundance_ratios}
    \begin{tabular}{ccc}
    \hline
    \hline
    Element &$\rm [X/Fe]$   &$\rm [X/O]$\\
    \hline
    C       &\CFe           &\CO        \\
    N       &\NFe           &\NO        \\
    O       &\OFe           &\nodata    \\
    Al      &\AlFe          &\AlO       \\
    Si      &\SiFe          &\SiO       \\
    S       &\SFe           &\SO        \\
    Fe      &\nodata        &\FeO       \\
    \hline
    \end{tabular}
\end{table}

%\begin{table}[htb]
%\centering
%\caption{Median [X/O] of VMP DLAs} \label{tab:med_XO}
%    \begin{tabular}{ccc}
%    \hline
%    \hline
%    Element &$\overline{\rm [X/O]}$    &2$\sigma$  \\
%    \hline
%    C       &\nodata    &err        \\
%    N       &\nodata    &err        \\
%    O       &0          &\nodata    \\
%    Al      &\nodata    &err        \\
%    Si      &\nodata    &err        \\
%    S       &\nodata    &err        \\
%    Fe      &\nodata    &err        \\
%    \hline
%    \end{tabular}
%\end{table}

%We plot the abundance ratios of the VMP DLAs with respect to Silicon in Figure \ref{fig:XSi_SiH}.

%\begin{figure*}
%    \centering
%    \includegraphics[scale=0.5]{Silicon/XSivsSiH_published_results_v5.pdf}
%    \caption{C, N, O, Al, Fe, and S abundance ratios as a function of [O/H]. Same symbols as Figure \ref{fig:XFe_FeH}.}
%    \label{fig:XSi_SiH}
%\end{figure*}

%\begin{table}[htb]
%\centering
%\caption{Median [X/Si] of VMP DLAs} \label{tab:med_XO}
%    \begin{tabular}{ccc}
%    \hline
%    \hline
%    Element &$\overline{\rm [X/Si]}$    &2$\sigma$  \\
%    \hline
%    C       &\nodata    &err        \\
%    N       &\nodata    &err        \\
%    O       &\nodata    &err        \\
%    Al      &\nodata    &err        \\
%    Si      &\nodata    &err        \\
%    S       &\nodata    &err        \\
%    Fe      &0          &0          \\
%    \hline
%    \end{tabular}
%\end{table}

\citet[][hereafter K19]{kirby+2019} placed empirical constraints on CCSNe using a method similar to what is presented here, except based on metal-poor stars in dwarf galaxies. We compare the empirical yields they derived to our empirical yields in Section \ref{sec:compare_yields}.

\subsection{Comparison Between VMP DLAs and VMP Stars} \label{sec:dla_stars_comparison}
We compare the abundance ratios of VMP DLAs to VMP stars in order to understand the robustness of these ratios. The grey background points in Figures \ref{fig:XFe_FeH} and \ref{fig:XO_OH} are metal-poor stars compiled from the JINAbase database with $-4 < [Fe/H] < -2$ from the MW halo, MW bulge, \edit1{ultra-faint dwarf (UFD)} galaxies, and classical dwarfs \citep{abohalima+2018}. The x's in the C and N panels are stars with surface gravity ($\log{g}$) small enough to necessitate corrections for evolution of surface abundances on the giant branch. All of the stellar abundances are subject to some nLTE correction, but the corrections for Al are particularly large.  The grey arrows in the figures show, to scale, the typical corrections needed to properly infer the abundance ratios;\textbf{\textnormal{ [C/Fe]$\sim$+0.5 dex \citep[][]{placco+2014} for $\log{g}<$1.6, [N/Fe]$\sim$-1 dex \citep[][]{placco+2014} for $\log{g}<$3.6, and [Al/Fe]$\sim$+0.8 dex \citep[][]{nordlander+2017} for stars with $-4 < [Fe/H] < -2$ and ${\rm [Al/Fe]} \sim -1$.}} The exact values of the corrections depend on temperature, surface gravity, metallicity, and for C and N, detailed abundances.

The red background points are a compilation of metal poor giants in \edit1{ultra-faint dwarf and classical dwarf (UF/D)} galaxies compiled by Alexander Ji\footnote{\dataset[https://github.com/alexji/alexmods/data/]{https://github.com/alexji/alexmods/blob/master/alexmods/data/abundance_tables/dwarf_lit_all.tab}} (references in Figure \ref{fig:XFe_FeH}).

Figure \ref{fig:XFe_FeH} shows that the VMP DLA abundance ratios for N and Al (i.e., [N/Fe] and [Al/Fe]) exhibit a distinct difference (up to $\sim 1$ dex) from VMP stars, whereas for the others ([C/Fe], [O/Fe], and [Si/Fe]) show general agreement between the samples. [Al/Fe] specifically show a systematic offset of -0.5 and can be explained by the nLTE corrections needed to infer their abundances. \edit1{Importantly}, measuring abundances from the cool dense gas from the DLAs is less susceptible to systematic effects than measuring abundances in stellar atmospheres. Sulfur is under-represented in the stellar sample because it has very few optical absorption lines. %Similar trends are seen in Figure \ref{fig:XO_OH}.

Figure \ref{fig:XO_OH} shows similar trends as the previous figure. There is a systemic offset between the VMP stars and VMP DLAs in [N/O] and [Al/O]. [C/O], [Fe/O], [Si/O] show general agreement between the samples.

%Some remarks on the agreement or disagreement between DLAs and stars.  Some insults to stars (LOL) and rough words that vindicate our choice to use DLAs by their superior moral virtue.

\edit1{Taken together, Figures \ref{fig:XFe_FeH} and \ref{fig:XO_OH} give another quantitative, visual affirmation for using VMP DLAs as a complementary set to VMP stars to constrain the yields of zero- and low-metallicity CCSNe \citep[e.g.,][]{prochaska+2002a,rafelski+2012,cooke+2017}.} For C, O, and Si, the abundances are complementary, whereas for N, Al, and S, VMP DLAs offer the ability to constrain yields without the effects of stellar astration (N), nLTE effects (Al), or observationally challenging wavelengths (S).

The empirical CCSN yields in Table \ref{tab:med_abundance_ratios} can be used as inputs in galactic chemical evolution models. Specifically, they could be an empirical guide to the first enriching events of the system being modeled.

\section{Theoretical Yield Comparison} \label{sec:theory_comparison}
In this section we compare the zero- and low-metallicity yields calculated by \citet[][hereafter WW95]{woosley+1995}; \citet{kobayashi+2006} and \citet[][together referenced hereafter as KN06]{nomoto+2006}; \citet[][hereafer HW10]{heger+2010}; \citet[][hereafter LC18]{limongi+2018}; and \citet[][hereafter \edit1{EC20}]{ebinger+2020} to the empirical yields we derived in the previous section.

%\subsection{Woosley \& Weaver 1995} \label{sec:ww95}

\subsection{Synopsis of the Theoretical Yields}
We discuss the relevant aspects of the physical models and input parameters that would affect the resultant yield predictions for the models below. For detailed discussions of each model we refer the reader to the cited manuscripts.

\subsubsection{\citet{woosley+1995}} \label{sec:ww95_discuss}
WW95 calculated the nucleosynthetic yields from massive stars as a function of mass and metallicity for elements through Zn. They computed 78 models that differed in explosion energy (usually $1.2 \times 10^{51}$~erg (1.2 B\@); B\@ is a Bethe, or $10^{51}$~erg.), metallicity (0--1 solar), and initial progenitor mass (11--40 $M_\odot$). We use the `Z' models, which have zero-metallicity, e.g., a pre-supernova progenitor with a Big Bang composition.

WW95 evolved each star through the supernova explosion.  The explosion was achieved by means of a mass piston located at the edge of the iron core modeled in 1-D\@ (modeling in 1-D\@ is common to all the models hereafter). The piston was moved inward at constant acceleration until it reached 500 km where it was moved rapidly outward (bounce) at a velocity tuned such that the final kinetic energy of the ejecta typically reached an energy of $\sim 10^{51} $ erg (1 B\@). The resultant shock decelerated in the mantle of massive stars, which, among other effects, led to significant fallback of Fe-peak elements. WW95's nucleosynthesis yields account for this fallback.

The models included the effects of neutrino irradiation, which had a major effect on the nucleosynthesis due to changes in the composition of the star before the shock wave from the mass piston caught up to the material.  Neglected processes that could have affected the nucleosynthesis include neutrino capture processes, stellar rotation, and a model of the explosion physics more realistic than a mass piston.  

The piston was tuned to ensure an explosion. Some modern 3-D supernova models explode without the need for a piston or thermal bomb by modeling the collapse and explosion phase of the star \citep[e.g.,][]{Burrows2019}, though there still is no consensus in the field as to a preferred method of explosion. %(though the authors add a "Mixing" term that is supposed to mimic the effect from the  stellar rotation).

We explore the effects that an ``explosion landscape'' (wherein some supernovae collapse into black holes without contributing to nucleosynthesis) on IMF-averaged theoretical yields in Section \ref{sec:compare_explosion}. Also, we use the WW95 yields as a qualitative reference only and perform our analysis (i.e., compare theoretical abundance ratios to VMP DLA abundance ratios) to their successor (see Section \ref{sec:hw10_discuss}.)

%\begin{figure}
    %\centering
    %\includegraphics[scale=0.6]{Orcid-ID.png}
    %\caption{The IMF-averaged CCSNe yield predictions from W95 compared to predictions %using our VMP DLA constrained yields. The theoretical yields are shown as orange %squares and the VMP DLA constrained yields are shown as blue octagons. O is %identical by construction.}
%    \label{fig:yield_comparison_w95}
%\end{figure}

%\subsection{Kobayashi et al. (2006) and Nomoto et al. (2006)} \label{sec:kn06}
\subsubsection{\citet{kobayashi+2006} and \citet{nomoto+2006}} \label{sec:kn06_discuss}
KN06 calculated the nucleosynthetic yields for elements up to Zn as a function of initial mass, composition, and explosion energy. We use their zero-metallicity yields over their total progenitor mass range (13--40 M$_\odot$) and compare their two available explosion energies: (1) $1 \times 10^{51}$~erg (1 B\@), corresponding to a normal Type~II supernova and (2) $10 - 30 \times 10^{51}$~erg (10--30 B\@), corresponding to a hypernova (HN). The authors found that a HN contribution of 50\% was optimal to match the observed [$\alpha$/Fe] trends in the Milky Way, but we vary the contribution from 0--100\% for our computed IMF-averaged yields (see Section \ref{sec:compare_yields}).

KN06 evolved the star from pre-supernova to explosion. Their explosion mechanism is a ``thermal bomb''; when a critical density is reached in the core, the star is promptly exploded with the specified energy for SNe and HNe. The authors used light curves and spectral fitting from individual SNe to set the mass of $^{56}$Ni ejected, 0.3--0.5 M$_\odot$. They also used the abundance ratios observed in EMP stars, specifically [O/Fe]=0.5, to set the amounts of mixing and fallback, which are otherwise free parameters in the calculations.

Included in their calculations were metallicity-dependent mass loss and neutron-capture processes.  They did not include any neutrino processes, stellar rotation, or natural explosion physics (as opposed to a thermal bomb).

%[[ If I remember correctly, Alex Ji told us that KN06 (and others?) calibrated their yields to observed stellar abundances.  This would be worthwhile to say here. ]]

%\subsection{Heger \& Woosley 2010} \label{sec:hw10}
\subsubsection{\citet{heger+2010}} \label{sec:hw10_discuss}
HW10 computed nucleosynthetic yields for zero-metallicity stars as a function of mixing (0.0-0.25; explained in the next paragraph), explosion energy ($0.1-10~$ erg; 0.1-10 B\@), mass cut (S4 or $Y_e$; also explained in the next paragraph), and progenitor mass (10--100 M$_\odot$). Several model combinations matched the abundances observed in different EMP stars adequately. One such model, as an example, had an explosion energy of 1.2 B\@, standard mixing of 0.1, and an S4 mass cut. These models are the successors to zero-metallicity models of WW95.

HW10 modeled their stars from main sequence to explosion using the KEPLER code, which was also used by WW95. They found that the density and structure of the zero-metallicity stars were the same as solar metallicity stars, implying a common central engine for single stars. The mixing of heavy elements deep in the star to outer layers was simulated by series of boxcar runs whose mass is set to be some fraction (0.0--25.1\%) of the He core mass. A mass piston was then used to explode the stars. There were two locations that the piston was placed: (1) near the base of the oxygen shell (where the entropy per baryon is $S/N_A$ = 4.0~$k_B$; S4 model) or (2) deeper in the star near the edge of the iron core (where the electron fraction $Y_e$ becomes discontinuous; `$Y_e$' model). Following this mass cut, each mass piston was moved to give the ejecta a final kinetic energy ranging from 0.3--10 $\times 10^{51}$ erg (0.3--10 B\@). 

The models include neutrino irradiation and fallback.  Neglected physics included stellar rotation (though the authors added a ``mixing'' term that mimics the effect of stellar rotation), neutrino winds (which would affect $r$- and some $s$-process elements), and more realistic explosion physics (as opposed to a piston). %There are certain regimes where stars either explode, fully implode, or fallback of metals synthesized explosively. 

\begin{figure*}
    \centering
    \includegraphics[scale=0.58]{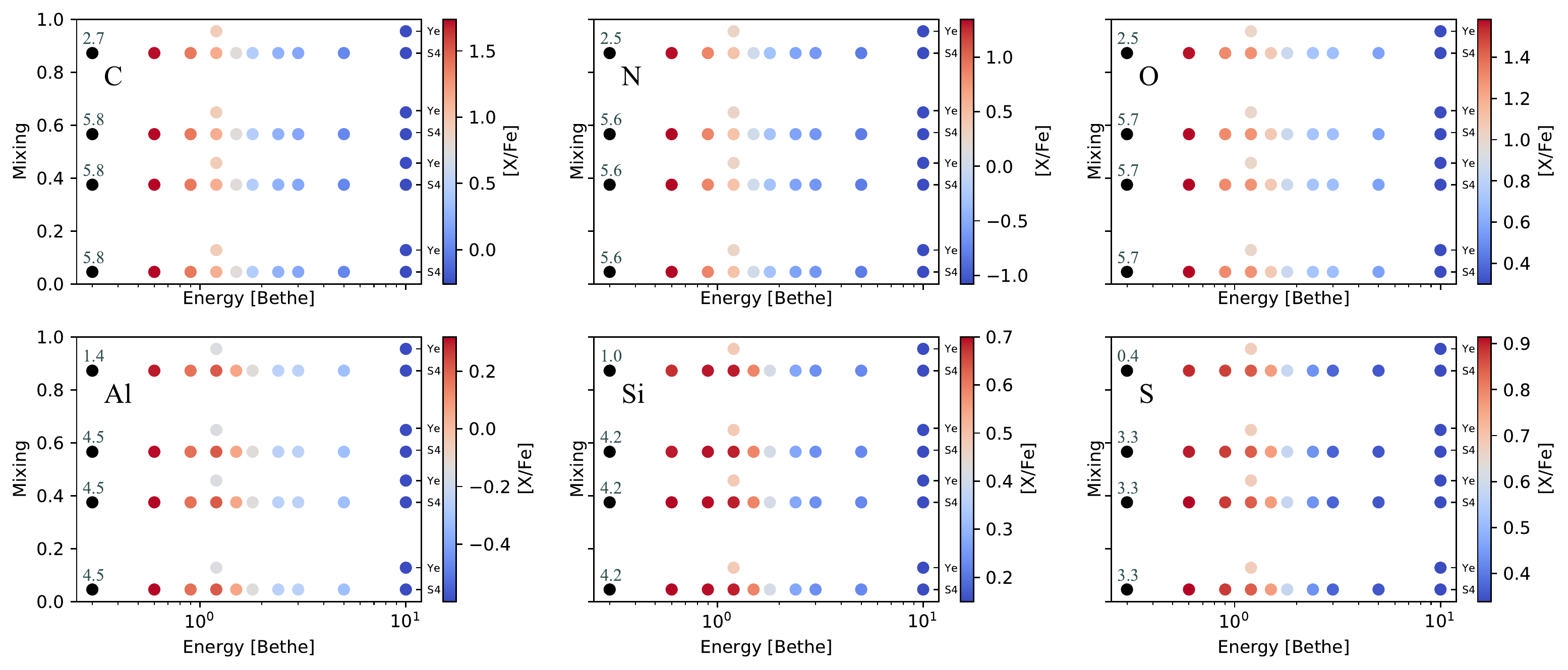}
    \caption{\textbf{\textnormal{The IMF-averaged yields from HW10 as a function of mixing parameter and explosion energy. The color bar on the right of the figure denotes the [X/Fe] for each mass cut, which are labeled on the right side of each panel. The $Y_e$ values are vertically offset from the $S4$ mass cut for clarity. The lowest explosion energy (0.3 B) points have their [X/Fe] value annotated above the point; they are all off the color bar scale.}}}
    \label{fig:hw10_yield_map}
\end{figure*}
 
To see how the numerous parameters affect the resultant yields, and to pick a preferred mass cut (S4 or $Y_e$) and mixing value (0.0--0.25), we show in Figure \ref{fig:hw10_yield_map} the IMF-averaged (Salpeter) yields as a function of mixing, explosion energy, and mass cut for the entire modeled mass range.  The yields stay constant with mixing but vary significantly ($\ge 3.5$ dex) with explosion energy. The insensitivity to mixing is expected because the mixing is performed after all explosive nucleosynthesis and fallback are computed. Because the $Y_e$ mass cut has only two explosion energies, its predicted [X/Fe] range is more tightly constrained than the S4 cut. Even so, the $Y_e$ range is always within the wider values predicted by the S4 cut so we elect to use the S4 cut in our comparison. The yields from the lowest modeled energy in the S4 models (0.3 B\@) differ significantly ($1-3$ dex) from the higher energy models, so we neglect the low-energy models in our comparison. After this culling, the yields are comparable regardless of mixing, so we choose a mixing value of 0, which leaves explosion energy and mass as the only free parameters in our comparison.

%\subsection{Limongi \& Chieffi et al. (2018)} \label{sec:lc18}
\subsubsection{\citet{limongi+2018}} \label{sec:lc18_discuss}
LC18 calculated the yields of elements up to Bi as a function of metallicity ([Fe/H] = $-3$, $-2$, $-1$, 0), progenitor mass (13--120~M$_\odot$), and initial rotational velocity ($v = 0$, 150, 300~km~s$^{-1}$). We use their ${\rm [Fe/H]} = -3$ model and masses from 13-100~M$_\odot$.  We discuss the choice of rotation velocity below.

LC18 modeled the evolution of their stars from pre-main sequence to presupernova. The initial metallicity given to the stars for their ${\rm [Fe/H]} = -3$ models were scaled from the solar composition \citep{asplund+2009} for most elements (e.g., Al, N) but abundance ratios (with respect to Fe) observed in metal-poor stars were used for some elements (C, O, Si, and S). They exploded their star via a thermal bomb with three different calibrations. Their preferred calibration (the one used in this work) requires that (1) stars from 13--25 $M_\odot$ have a mixing and fallback scheme by requiring that the edges of the mixing region are fixed and that the mass cut is placed such that $0.07~M_\odot$ of $^{56}$Ni is produced and (2) stars more massive than 25 $M_\odot$ fully implode and therefore any yields are from the pre-SN stellar wind. The explosions energies that the authors calculated, which we infer from their quoted binding energy of mass above the Fe core, ranged from 0.65--15 $\times 10^{51}$ erg (0.65--15 B\@) for their 13--80 M$_\odot$ progenitors with the larger explosion energies corresponding to the more massive progenitors. We fully decayed the isotopes in our yield comparison to their final stable isotopes assuming 100\% conversion.

LC18 included stellar rotation and mass loss but neglected realistic explosion physics (as opposed to a thermal bomb).

Stars are known to have varying rotation velocities so there likely exists some preferred Initial Distribution of ROtation Velocities (IDROV; analogous to mass and the IMF), that is a function of mass and metallicity \citep[][]{limongi+2018,prantzos_2018}. At present, there has not been a detailed study of the IDROV and its properties coupled with the IMF. Still, we construct a simple, observationally motivated IDROV (see Section \ref{sec:compare_explosion}), alongside the yield predictions when all stars are rotating at a single velocity (see Section \ref{sec:compare_yields}). %But given the few observational constraints for the IDROV, and its sensitivity to the IMF, we do not adopt an IRDOV but instead show the velocities separately.
%IMF, we omit this and show the velocities separately (\textbf{REFERENCES}).

%\subsection{Ebinger et al. (2020)} \label{sec:pe20}
\subsubsection{\citet{ebinger+2020}} \label{sec:pe20_discuss}
\edit1{EC20} of the \edit1{PUSH collaboration \citep[][]{perego+2015}} used an engine to self-consistently explode stars and obtain nucleosynthetic yields for isotopes up to $^{211}$Eu as a function of metallicity ($Z=0-10^{-4}\rm \; {Z}_\odot$) and mass (11-75 $M_\odot$). We use their zero-metallicity models, which have a smaller mass range of 11--40 M$_\odot$. 

The authors used pre-SN models from \citet{woosley+2002}. The explosion, which relies on the delayed neutrino-driven mechanism, of the pre-SN progenitor was simulated by following the core collapse, bounce, neutrino heating, and resultant explosion (or implosion) of the stars in increments of 1 M$_\odot$ assuming spherical symmetry. The neutrino heating term, the most important term for the explosion, has two free parameters that were calibrated such that the models reproduced properties observed from SN 1987A\@. 

 The authors did not force the stars to explode but instead relied on the physical outputs from their simulation to determine if an explosion was successful or not. An explosion was deemed successful if the final explosion energy of the simulation was positive. An explosion was deemed as a failure if the final explosion energy was negative at the end of the simulation. From this criteria they found that stars from 11--23 M$_\odot$ and 27--31 $M_\odot$ successfully exploded whereas the other masses failed i.e., imploded into a black hole. The exploding stars contributed to yields whereas direct-collapse stars did not. The resultant explosion energies ranged from $\sim$0.3--1.6$\times 10^{51}$ erg (0.3--1.6 B\@) with no uniform trend with progenitor mass.

\edit1{Only explosive nucleosynthesis was considered in the yields published by EC20, so we added the contribution from the preSN progenitors (i.e., hydrostatic nucleosynthesis) to get the total yield \citep[S. Curtis, private communication,][]{woosley+2002}. This primarily affected the abundance of light elements (C, N, O) that are created in small quantities during explosive nucleosynthesis. We could not compute a total yield for Al because the pre-SN progenitors did not track it. We do not include the Al predictions from EC20 in our subsequent analysis.}

The authors included realistic explosion physics but neglected neutrino irradition and stellar rotation.

%[[ here and in previous subsubsections: discussion of which mass range is presumed to explode and therefore produce yields ]]

\subsection{Comparison Between Yield Tables} \label{sec:compare_yields}
\begin{figure*}
    \centering
    \includegraphics[scale=0.7]{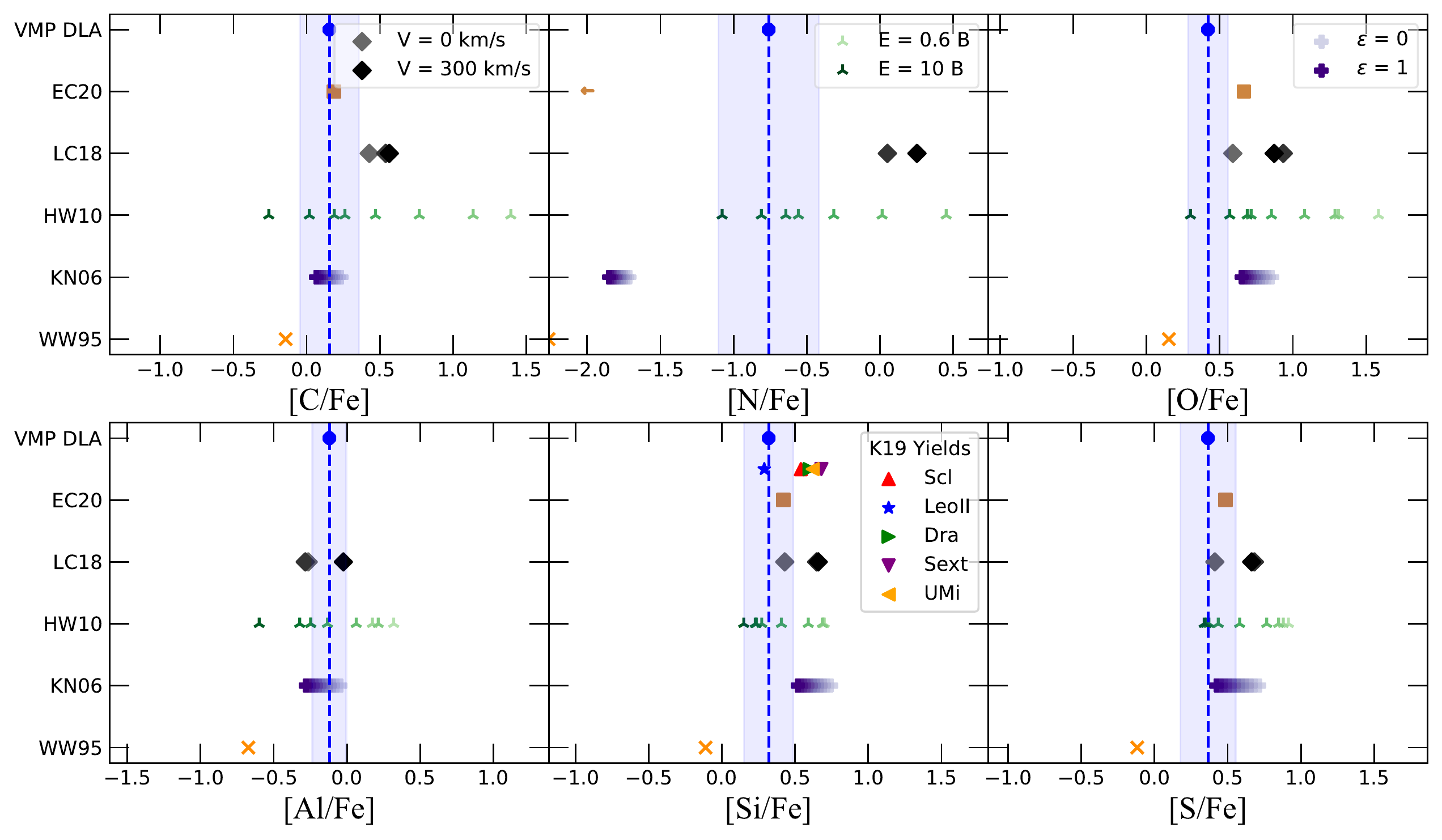}
    \caption{\edit1{Predicted [X/Fe] for each yield table compared with the median [X/Fe] from VMP DLAs. The dashed vertical blue line is the median of the VMP DLAs, the blue shaded region shows the 1-$\sigma$ uncertainties. \edit1{The colored stars are empirical yields derived by K19 from metal-poor stars in the respective dwarf galaxies (see end of Section \ref{sec:empricial_yields}).} \edit1{EC20} predictions are shown as orange squares, LC18 as black/dark grey diamonds (varying rotational velocities), HW10 as green tripoints (varying explosion energy), KN06 as blue pluses (varying HN contribution), and WW95 as orange X's.}}
    \label{fig:yield_comparison_Fe}
\end{figure*}

In Figure \ref{fig:yield_comparison_Fe} we show the theoretical [X/Fe] predictions from the models described in the previous section and compare them to empirical yields from VMP DLAs (see Section \ref{sec:empricial_yields}). The [X/Fe] values were calculated by integrating the model yields under a Salpeter IMF from a mass range of 10--100 $M_\odot$. We interpolate the yields to achieve steps of 0.25 $M_\odot$ but did not extrapolate the yields outside of their modeled masses (see mass ranges above). For [Si/Fe] we also compare the empirical yields derived by K19 using metal-poor stars in the dwarf galaxies Sculptor (Scl), Leo II, Draco (Dra), Sextans (Sext), and Ursa Minor (UMi) \edit1{(see end of Section \ref{sec:empricial_yields})}.

There is general agreement between the empirical yields and the theoretical [X/Fe] predictions for the majority of elements. Each model reproduces the empirical yields (within $1-\sigma$) for a subset of elements. [Si/Fe] and [S/Fe] are consistently reproduced by all models. [C/Fe], [O/Fe], and [Al/Fe] show similar agreement \edit1{(\edit1{EC20} did not track the total Al yields; see Section \ref{sec:pe20_discuss})}. [N/Fe] is a notable exception, it showed the largest spread between models. Only one, HW10, reproduced the empirical [N/Fe] yields. The [N/Fe] differences between the empirical yields and theoretical predictions exceeded 0.5 dex for all other models.

\edit1{The various models' overall agreement in their [S/Fe] and [Si/Fe] predictions with the VMP DLA data can be used to further argue that there is minimal dust depletion in VMP DLAs. If this were not true, one would expect systematic discrepancies between the nucleosynthetic predictions and the observed ratios in VMP DLAs, just as one would expect some trend in [Si/Fe] and [S/Fe]. But neither is seen, further suggesting that there is negligible dust depletion in VMP DLAs.}

The majority of the yields derived by K19 fall just outside of our uncertainties, except for Leo II\@. After accounting for K19's yield uncertainties (0.02--0.08 dex), Scl and Dra also become consistent with the DLA estimates. Note that K19 measured Si in red giants in a spectral range $\sim$6300--9100~\AA\@.  The only available Si lines in this spectral range are weak due to their high excitation potentials, making it difficult to measure Si in low-S/N spectra. This means there could be a bias toward higher Si abundances for very metal-poor stars because lower Si abundances would result in undetectably weak lines. In fact, extremely-metal poor Milky Way halo stars have a plateau at ${\rm [Si/Fe]} = 0.4$ \citep[e.g.,][]{Frebel2015}, which is consistent with the VMP DLA abundances.

\begin{figure*}
    \centering
    \includegraphics[scale=0.7]{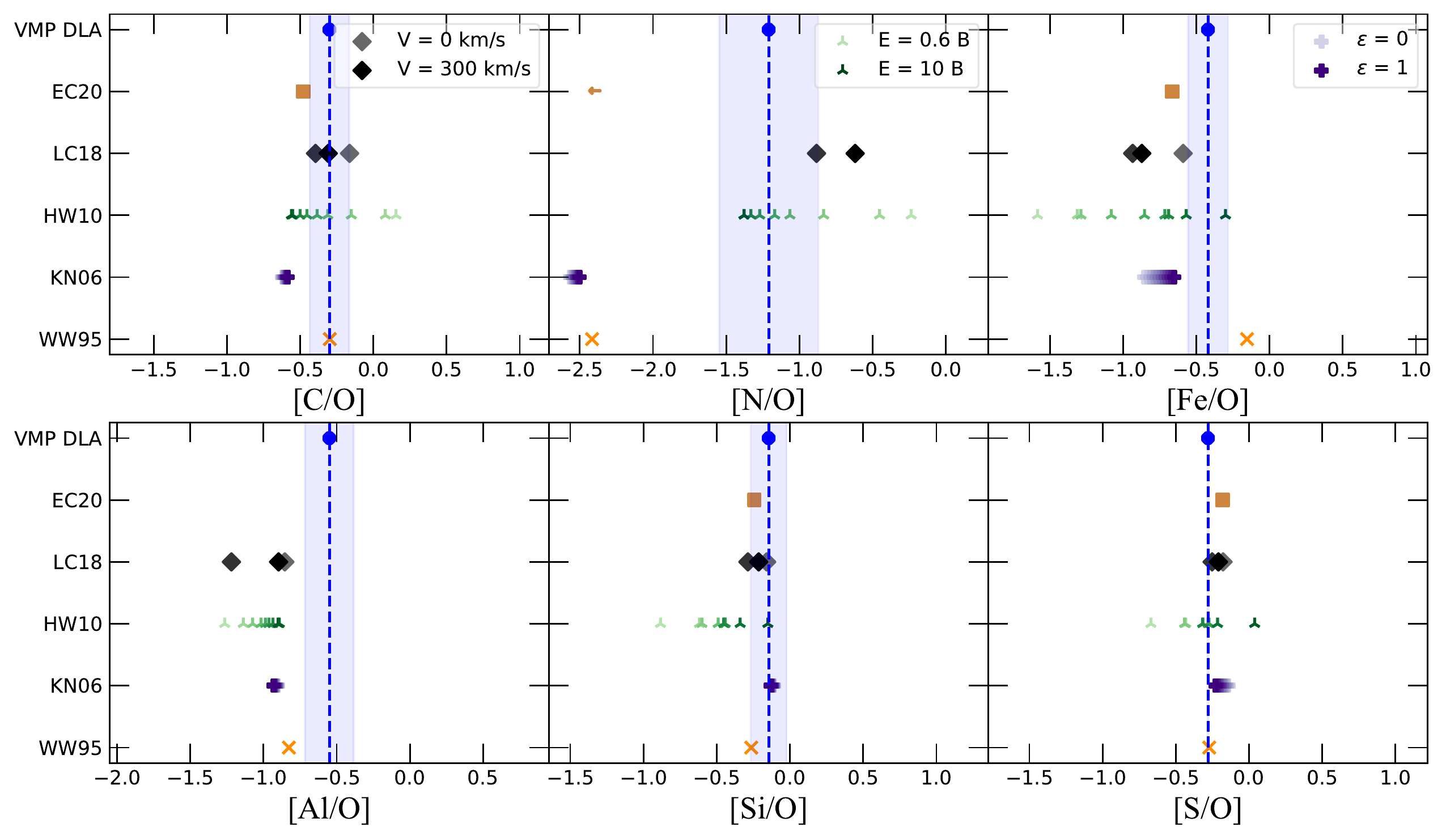}
    \caption{\edit1{Median [X/O] for VMP DLAs and predicted [X/O] for each of the yield tables. Same symbols as Figure \ref{fig:yield_comparison_Fe}.}}
    \label{fig:yield_comparison_Oxygen}
\end{figure*}

In Figure \ref{fig:yield_comparison_Oxygen} we compare the theoretical [X/O] to the empirical yields. The theoretical [X/O] values were calculated using the same method as above for [X/Fe].

There is general agreement between the inferred empirical yields and the models for a subset of elements, Si and S. \edit1{We see near unanimous agreement between models in their predictions for [C/O], [Fe/O], [Si/O], and [S/O].}%, except for \edit1{EC20}.  
(There is only one VMP DLA data point available for [S/O].) Wide disagreement is seen in [N/O] where only two models reproduced the empirical yields, HW10 and LC18. No model reproduces [Al/O]. %The disagreement is more apparent for [N/O] though as the aluminum abundance ratios are all at the border of the 1-$\sigma$ uncertainties.

%The wide disagreement seen in both [N/Fe] and [N/O] point to a common overabundance or deficiency of N predicted by the majority of models (except HW10). In other words, the models seem not to produce the amount of N that the empirical yields suggest. The consistency seen in [Al/Fe] and [Fe/O] (therefore [O/Fe])  makes the wide disagreement seen in [Al/O] unexpected. Note that the disagreement seen is $\leq$ 0.1 dex for almost all models (except PE20). 

%In Figure \ref{fig:yield_comparison_Oxygen} we plot the [X/Si] predicted from the theoretical yield tables and the median values from the VMP DLAs.

%\begin{figure*}
%    \centering
%    \includegraphics[scale=0.5]{Silicon/comparison_plot_Silicon_no_extrap_v3.pdf}
%    \caption{Median [X/Si] for VMP DLAs and predicted [X/Si] for each of the yield table. Same symbols as Figure \ref{fig:yield_comparison_Fe}.}
%    \label{fig:yield_comparison_Silicon}
%\end{figure*}

\subsection{Explosion Landscape and IDROV} \label{sec:compare_explosion}
 %In this section we investigate the effects of adopting an explosion landscape or an Initial Distribution Rotation of Velocities (IDROV) to the HW10 and LC18. yields respectively.
 All of the models discussed thus far had their parameters tuned to ensure that the star exploded, except \edit1{EC20}. This was also true for LC18, though they ensured that stars above 25 M$_\odot$ fully imploded. There is strong evidence that massive stars in certain regions of mass--metallicity space will collapse directly to a black hole, resulting in zero metal yield, other than any yields from pre-SN winds \citep{adams+2016,smartt+2015,sukhbold+2016,woosley+2017}. This so called SN explosion landscape describes the ability of stars to explode or not by assigning one of \edit1{two} outcomes to ranges of stellar mass (at fixed metallicity): \edit1{(1) explosion potentially with fallback or (2)} direct collapse into a black hole. \edit1{Fallback onto the dense remnant during/after explosion has been shown to be rare in recent studies \citep{sukhbold+2016,ertl+2020}. Still, given \textbf{\textnormal{it's}} importance, all of the models we described previously (Sections \ref{sec:ww95_discuss}--\ref{sec:pe20_discuss}) include fallback in their calculations. %Therefore, we technically do not (cannot) deal with case (1) in our analysis. %, though HW10 implicitly include fallback in their yields. 
 We add the yield contributions for exploding stars (i.e., the yields with fallback included) and remove 100\% of the yields from the imploding stars.} % and assume stars either fully explode or implode. 
 
 Because there has not been a recent SN explosion landscape for zero- or low-metallicity stars that accounted for the full mass range (10--100 $M_\odot$), we use results computed by \edit1{EC20} for the mass range  %three authors to do so; 
 10--40 $M_\odot$. \edit1{T}hey found that stars exploded in the mass ranges 11--23 and 27--31 $M_\odot$ (in steps of 1 $M_\odot$) but did not explode otherwise (see Section \ref{sec:pe20_discuss} for discussion of the \edit1{EC20} models). We summarize the explosion landscapes from \edit1{EC20} and LC18 in Table \ref{tab:explosion_landscape}.
 %, 40--70 M$_\odot$\citep[][hereafter S16]{sukhbold+2016}, 70--100 M$_\odot$ \citep[][hereafter W17]{woosley+2017}, as summarized in Table \ref{tab:explosion_landscape}. 

 \begin{table}[htb]
\centering
\caption{SN Explosion Landscapes} \label{tab:explosion_landscape}
    \begin{tabular}{cccc}
    \hline
    \hline
    Model Range  &Step      &Explosion   &Authors\\
    (M$_\odot$) &(M$_\odot$)                &(M$_\odot$)    &       \\
    \hline
    10--40      &1              &11--23; 27--31     &\edit1{EC20}\\
    13--120     &2, 5, 10,  20              &13--25             &LC18\\
    \hline
    \end{tabular}
\end{table}

\begin{figure}
     \centering
     \includegraphics[scale=0.65]{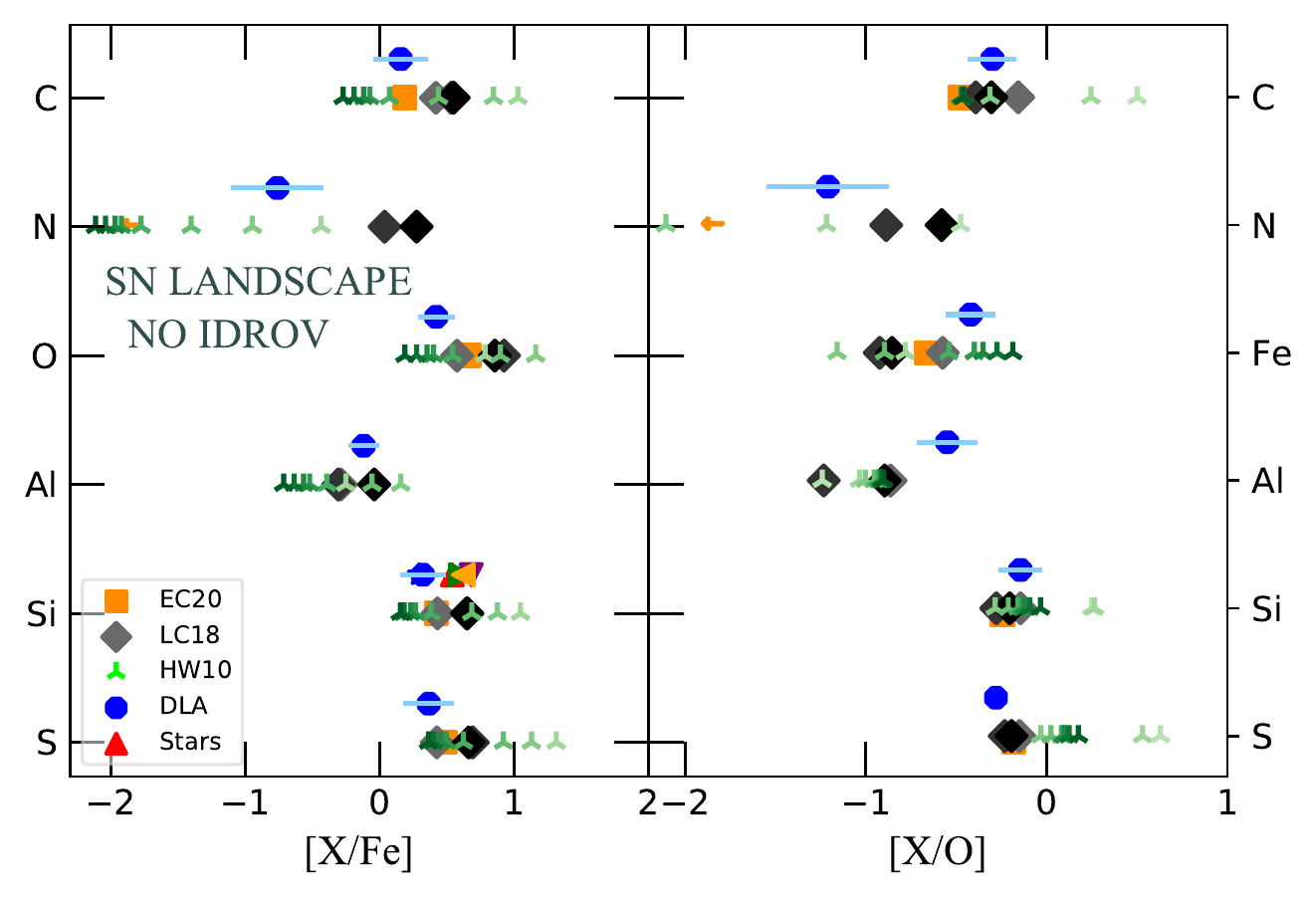}
     \includegraphics[scale=0.65]{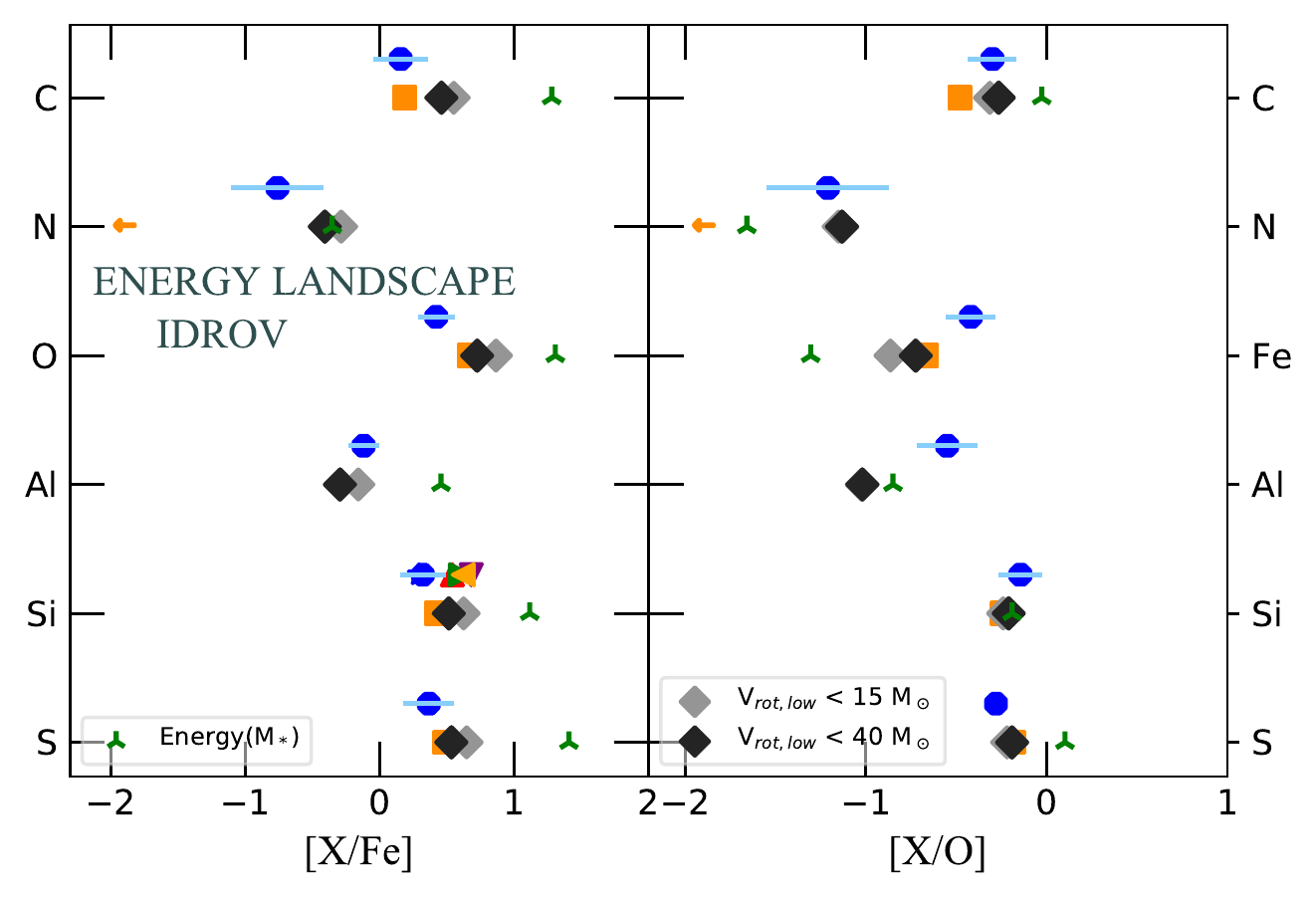}
     \caption{\edit1{Theoretical yield ratios compared to the empirical abundance ratios\textbf{\textnormal{\; [X/Fe] on the \textit{left} and [X/O] on the \textit{right}}}. Same symbols as Figure \ref{fig:yield_comparison_Fe} unless noted otherwise. The empirical yields and uncertainties (from DLAs and stars \textbf{\textnormal{([Si/Fe])}}) are vertically offset for clarity.
     \textit{Top:} Predicted yields from HW10 when imposing a SN explosion landscape, and as-published LC18 and \edit1{EC20} yields.
     \textit{Bottom:} HW10 yields when imposing both the explosion landscape and explosion energy landscape (i.e., fixing explosion energy), and LC18 yields when imposing an IDROV such that all stars up to a given mass (e.g., 15 M$_\odot$; light grey diamond) rotate with a low initial rotation velocity (50 km s$^{-1}$) while stars more massive rotate at 250 km s$^{-1}$ (dark grey diamond).}}
     \label{fig:explosion_physics}
 \end{figure}

 To understand how the inclusion of realistic explosion physics changes the predicted yields, we show in the top panel of Figure \ref{fig:explosion_physics} abundance ratios from HW10 modified by the SN explosion landscape from \edit1{EC20} (i.e., only adding the yields from stars that the authors found to explode and removing the yields from those that do not) and as-published LC18 and \edit1{EC20} yields, along with the empirical DLA abundances. We chose not to adapt this SN explosion landscape to the other yield models because they used different stellar evolutionary codes. The landscape is sensitive to the final core structure of the star, which itself is sensitive to all prior modeling assumptions (e.g., stellar evolutionary code, interaction cross sections, adopted reactions rates, convection criteria, etc.) such that two authors using different assumptions will likely have different landscapes. The progenitors used in HW10 and \edit1{EC20} used the same formalism, based on the KEPLER code for stellar evolution, as were the the progenitors from \citet{woosley+1995} and \citet{woosley+2002}.  Even though there are differences present between the the progenitors of HW10 and \edit1{EC20}, they are closer to direct comparisons than the other models, who used their own stellar evolutionary code.%LC18 imposed their own landscape by requiring that all stars above 25 M$_\odot$ fully imploded, we therefore compare their yields in this section.
  We calculate the yields in the same manner as the previous section except that we do not interpolate between the modeled masses to ensure that only the stars that explode contribute to the yields. We compared this method with the interpolation method of Section \ref{sec:compare_yields} and found that they are comparable, with differences $< 0.01$ dex.

  %In order to , we surveyed work over the past few decades on the supernova explosion landscape. 
 %PE20 constructed an explosion landscape as a consequence of their modeling process that we impose on the yields from HW10.  We chose HW10 because the explosion landscape was designed from models very similar to them. Instead, we add 100\% of the yield contributions from stars that explode and remove 100\% of the yield contributions from stars that do not.
 
 %S16 determined the explosion landscape for stars of solar metallicity from 10 to 100 $M_\odot$ using a combination of observational constraints (e.g., remnant masses) and theoretical constraints. The authors model the explosion of the stars using 5 different central engines (in steps of 5 $M_\odot$ in the mass range 40--70 $M_\odot$). They find that the stars implode in all 5 central engines between 40--55 $M_\odot$ and 65--70 $M_\odot$, so we remove the yield contributions from stars in this mass range. For 3 out of 5 of their models, they find that the 60 M$_\odot$ star explodes so we include the yields for this mass.

 %W17 calculated the explosion landscape for very massive (70--150 $M_\odot$) zero- and low-metallicity stars in which PISN is the primary explosion mechanism for the stars. They found that all stars in this mass range (in steps of 5 M$_\odot$) exploded except the 70 $M_\odot$ model so we remove its yield contribution and keep the rest.
 
 Similar to the SN explosion landscape, one can analogously construct an explosion energy landscape that maps progenitor mass to explosion energy (i.e., remove explosion energy as a free parameter). \edit1{EC20} predicted explosion energies ranging from 0.3--1.6 B %that did not increase monotonically with mass but instead showed 
 with a peak of $\sim$1.6 B\@ for the 25 $M_\odot$ progenitor, and a minimum of 0.3 B\@ for the 31 M$_\odot$ progenitor. We adapt this explosion energy landscape to HW10 alongside their explosion landscape (i.e., we ensure that only stars that explode contribute to the yields, and that the energy for each progenitor reflects the energy calculated by \edit1{EC20} for the progenitor mass), and show the IMF-averaged yields in the bottom panel of Figure \ref{fig:explosion_physics}.
 
 \edit1{We discussed briefly} in Section \ref{sec:lc18_discuss}, that there likely exists a preferred IDROV which one can map from progenitor mass to rotation velocity. There is strong evidence for a bimodal velocity distribution among young massive stars in the local universe in which there are slow rotators (40--60 km s$^{-1}$) and fast rotators (150--300 km s$^{-1}$) \citep[][]{ramirez-agudelo+2015,milone+2018,kamann+2020}. Whether a massive star is a fast or slow rotator does not strongly depend on its mass or its membership in a binary system \citep[][]{bouvier+2013,bastian+2020,kamann+2020} .%; a binary association be used as a proxy for mass (i.e., more massive stars tend to be in binary configurations with comparable mass companions \citep[][]{kobulnicky+2007}). Though the general trend would seem to imply that more massive stars rotate more rapidly. 
 We construct a simple, observationally motivated IDROV that we apply to the LC18 yields in which we require that all stars below a mass cutoff (15, 25, 30 or 40 M$_\odot$) are slow rotators with $v_{rot,low}$ = 50 km s$^{-1}$, and stars above the cutoff are rapid rotators with $v_{rot,high}$ = 250 km s$^{-1}$. We show the IMF-averaged, IDROV-modified LC18 yields in the bottom panel of Figure \ref{fig:explosion_physics}. 
 
 This IDROV is based on observations of local, solar-metallicity, massive stars.  The IDROV might be different at low metallicity. Because there is little to no observational evidence on how rotational velocity or binarity changes at low metallicity \citep[][]{maxwell+2017}, we adopt the IDROV mentioned above with the acknowledgement that it will likely need to be modified when the necessary data is present.
 
The inclusion of the explosion landscape into HW10 does not make a significant difference in the predicted abundance ratios (less than 0.2 dex for most elements). The behavior seen here is similar to that in Figures \ref{fig:yield_comparison_Fe} and \ref{fig:yield_comparison_Oxygen}, where for every element HW10 has at least a few explosion energies (typically between 3--10 B) that fall within the range of the empirical constraints; LC18 reproduces most of the abundances; and \edit1{EC20} is well within the empirical estimates for some elements and differs by more than 1 dex for others. 

Imposing the explosion energy landscape onto the HW10 yields results in widespread disagreement with the empirical yields; more than half of the predicted abundance ratios fall well outside the empirical yield uncertainties. The largest disagreement is for [X/Fe], where only one empirical ratio is reproduced, [N/Fe]. Similarly for [X/O], only one ratio is reproduced, [Si/O].

The IDROV-modified LC18 yields nearly unanimously reproduce the empirical yields (except [Fe/O] and [Al/O]), an improvement compared to the simpler cases shown in the top panel of Figure \ref{fig:explosion_physics} (and Figures \ref{fig:yield_comparison_Fe} and \ref{fig:yield_comparison_Oxygen}) where all stars are assumed to rotate at the same velocity. The mass cutoffs in the range 25--40 M$_\odot$ predict comparable yields, whereas the 15 M$_\odot$ cut-off differed by a small amount ($<$ 0.2 dex).

\section{Lessons from VMP DLAs} \label{sec:discussion}
%Some preliminary conclusions on which bits of the input physics do a good job of reproducing the VMP DLA yields.

In this section we discuss the physical reasons that could explain why each set of yields matched (or did not match) the VMP DLA abundance ratios.

The large variation in explosion energy of HW10 ensured that at least one energy was able to reproduce the empirical yields for most abundance ratios (except [Al/O]) when no explosion landscape or energy landscape was adopted. The energy range that best fit the data varied between 3--10 B\@. Although we cannot place any constraints on a preferred mixing treatment (because they all predict comparable abundance ratios; see Figure \ref{fig:hw10_yield_map}), we focus on the S4 mass cut over the $Y_e$ mass cut because of the larger range of explosion energies that were modeled. KN06 modeled a similarly wide energy range (1--30 B), but their yields did not reproduce the empirical abundance ratios for as many ratios as HW10. One reason for this could be the calibrations used by KN06 of the amount of $^{56}$Ni that each star had to produce. This observational constraint has the effect of restricting the predicted abundance ratios to a smaller range of values than HW10, who did not impose such constraints in their models.
When we applied the SN explosion landscape calculated by \edit1{EC20} in two different ways onto HW10 we found 1) comparable yield predictions when energy is left as a free parameter, but 2) wide disagreement when energy is constrained to be a function of mass. Both points are consistent with the main HW10 result (without the landscape constraints), that higher-energy explosions ($\geq$ 3 B\@) were needed to reproduce the data. The disagreement in [X/Fe] with the fixed, lower energies can be explained by the resultant low Fe yield from these models, which systematically increased the [X/Fe] ratios for the light and intermediate mass elements (i.e., those synthesized hydrostatically).

LC18, similar to HW10, consistently reproduced the VMP DLA yields with at least one of their models (varying initial rotational velocity) falling within the empirical estimates for each abundance ratio except [N/Fe] and [Al/O]. Similar to KN06, LC18 tuned their models to reproduce observational constraints on the amount of $^{56}$Ni produced in the supernova. They achieved this tuning by varying the location of the mass cut for their pre-SN progenitor. %There was little discussion on the typical explosion energies the authors saw. %but they were likely in the range of 1--10 B as this is observationally favored and seems to consistently reproduce the empirical yields. 
The initial rotation velocity that reproduces the empirical yields %$\sim 0.3~\rm dex$ of the 1-$\sigma$ confidence interval of the empirical yields 
for most ratios was 0 km s$^{-1}$ (exception to this was [N/O], where the higher rotation velocities reproduced the data).
When a simple IDROV is adopted, in which stars below a certain mass threshold were considered slow rotators (50 km s$^{-1}$) and stars above that mass cut were considered fast rotators (250 km s$^{-1}$), we found near unanimous agreement between the predicted yields and the empirical yields \edit1{(exceptions were [Fe/O] and [Al/O]).} We acknowledge that this is a simple parameterization that could be improved with better data on the rotation velocities and binarity of low-metallicity massive stars, and/or when binary stellar evolution is modeled in more detail. Even so, our findings may suggest that adopting some observationally motivated IDROV is helpful in matching observed abundance ratios of VMP DLAs.

KN06 reproduced the empirical yields of most abundance ratios to within $\sim 0.1~{\rm dex}$ of the 1-$\sigma$ confidence interval. A HN contribution of $\gtrsim 50\%$ was required to reproduce more than half of the empirical yields. Exceptions to this agreement were [N/Fe], [N/O], [C/O], and [Al/O]), which were $\sim 0.3--1~{\rm dex}$ discrepant with the empirical yields. Because N is synthesized hydrostatically during H burning, and is therefore not sensitive to the explosion mechanism of the star, one explanation for this discrepancy could originate in the pre-SN evolution of the star. In contrast, there the Fe yield could also be the cause of the discrepancy.% \citep{woosley+1995,woosley+2002}. %Another explanation could be N falling back onto the dense remnant after the explosion of the star \edit1{Tuguldur, what do you think?}.

\edit1{EC20} reproduced the empirical yields for \edit1{more than half} of the ratios we studied \edit1{(C,} O, Si, and S with respect to O and Fe). \edit1{N  disagreed with the empirical yields by up to a few orders of magnitude (Figures \ref{fig:yield_comparison_Fe}, \ref{fig:yield_comparison_Oxygen}, and \ref{fig:explosion_physics}), and Al did not have a prediction for its total yield, so it was not included in our analysis.} \edit1{EC20} uniquely modeled the collapse and explosion phase of the SN for each of the stars they modeled, but they did not model the MS evolution of the stars and instead used pre-SN progenitors from the literature (see Section \ref{sec:pe20_discuss}). \edit1{Nitrogen, which is synthesized hydrostatically, showed the largest discrepancy between the EC20 yields and the DLAs. As \edit1{argued} previously, this discrepancy likely originates in the pre-SN evolution of the stars. Another (less plausible) explanation could be that the explosion landscape itself causes the discrepancy. Perhaps the combined mass of N that was lost to implosion, if allowed to contribute to the yields, could account for the differences between the predictions and the data (but see \citet[][]{griffith+2021b}).} %\edit1{Tuguldur, thoughts?}. 

\edit1{\edit1{EC20}'s predicted abundance ratios for the intermediate-mass and heavy elements studied here (S, Si, and Fe) consistently reproduced the empirical yields. This suggests that the explosion mechanism employed by the authors is consistent with the data.} %elements are more sensitive to the explosion mechanism of the star, so the intermediate-mass element agreement coupled with the light element disagreement, suggests that the explosion mechanism employed by the authors is consistent with the data whereas the pre-SN evolution is not.

There is an interesting discrepancy between HW10 (when the explosion energy landscape is applied) and \edit1{EC20}. One would expect agreement between the two models because they are based on similar progenitors, have the same explosion landscape, and have the same energy constraints. But the bottom panels of Figure \ref{fig:explosion_physics} show that there is still an average difference of 1 dex between them. Specifically, the \edit1{EC20 [X/Fe] ratios agree with the DLAs for most of the elements including the intermediate-mass elements ([Si/Fe] and [S/Fe])}.  In contrast, HW10 shows systematically high [X/Fe] for all ratios. \edit1{EC20 has low [N/O], but they reproduced the empirical [X/O] for all other elements,} contrasted by HW10 which only reproduced [Si/O]. \edit1{As mentioned previously, HW10 produced very little Fe in these low explosion energy models, and \edit1{EC20} conversely %, and interestingly, 
produced a lot of Fe in their lowest explosion energy models. The low iron yields from HW10 are explained by increased fallback of iron-group elements with decreasing explosion energy% where, interestingly, \edit1{EC20} see the opposite effect. %As mentioned previously, \edit1{EC20} seems to under-produce the yields of light element N. 
%Therefore, it seems
. Therefore, fallback treatment is likely causing the large discrepancy between the two models.}%That might be a gradient in the PE20 yields where more massive species seem to be more abundant than lighter elements. In other words, the differences between HW10 and PE20 may arise from the underproduction of light elements in PE20 which itself, is tied to the pre-SN progenitor more than the explosion mechanism of the models.

A common feature in the models that reproduced the data were high explosion energies. We showed in Figures \ref{fig:yield_comparison_Fe} and \ref{fig:yield_comparison_Oxygen} that the models that best reproduced the empirical yields were those with explosion energies exceeding 2 B\@; namely KN06, HW10, and LC18. For calibrated neutrino-driven explosions it is difficult to exceed this threshold, as seen in the peak energy explosion of 1.6 B\@ calculated by \edit1{EC20} \citep[See also][]{perego+2015,ertl+2016, sukhbold+2016, ertl+2020}. As discussed in Section \ref{sec:hw10_discuss}, it has been shown \citep{heger+2010} that the core structures for solar metallicity stars and metal-free stars are similar, so it should be the case that the explosion and explosive nucleosynthesis in these models should be similar.  Our work adds more evidence to suggest that the energetics of the explosion, and potentially the underlying explosion mechanism, must be modified to allow for higher energies if one is to match the abundances measured from VMP stars and VMP DLAs. \edit1{An interesting point of contention for this is the fact that EC20, who had comparatively low explosion energies, consistently reproduced the abundance ratios of half of the light elements (C and O) and all of intermediate-mass elements. This suggests that that there may be no need for HN class explosion energies if one properly models the explosion phases of the delayed neutrino-driven mechanism.}%This need for higher explosion energies has been observed before when trying to match the abundances of metal poor stars but may be the first time this is seen from VMP DLAs.

\edit1{W}e do not have much observational data on how binarity of massive stars changes at low-metallicity \citep[][]{maxwell+2017}, and are just starting to theoretically explore the effects of binarity on their evolution and final outcomes \citep[][]{ertl+2020, vartanyan_+2020}. A potential solution for the low explosion energies of neutrino-driven explosions would be to invoke a rotationally powered explosion \citep[][]{mosta+2015} from stars that have evolved in binary configurations where the progenitor gains a substantial rotational energy through its companion. These rotationally powered explosions could achieve explosion energy comparable to HN ($\sim10$ B).

\section{Conclusions} \label{sec:conclusions}
We have placed empirical constraints on the CCSN yields of zero- and low-metallicity stars using the abundances measured from VMP ([Fe/H] $<$ $-2$) DLAs available in the literature from the past 30 years.  The majority of this compilation is based on high-resolution spectroscopic measurements (Section \ref{sec:empricial_yields}). We equated the median of the VMP DLA abundances with the IMF-averaged CCSN yield by assuming that the VMP DLAs are at the earliest stages of galactic chemical evolution, where CCSNe dominate the nucleosynthesis.

\edit1{We show that our approach is complementary, and at times superior, to using VMP stars for the same work \citep[Section \ref{sec:dla_stars_comparison}; e.g.,][]{berg+2015,welsh+2019}.} In particular, for elements whose stellar photospheric abundances depend on the assumption of LTE (O, Al), astration corrections (N), or difficult-to-measure atomic transitions (O, S), VMP DLAs are superior because measuring abundances from their cool, mostly neutral gas relieves the need for such corrections or considerations. %For other elements where these corrections and considerations are negligible in stars (e.g. O, Si, and Fe) our work is complementary.

We compare the empirical yields to the most widely adopted theoretical yields in the literature and find that all models can reproduce the empirical yields for a subset of abundance ratios studied here by varying only a single parameter (e.g., explosion energy, HN contribution, or initial rotation velocity). The yields calculated by HW10 (Section \ref{sec:hw10_discuss}) consistently reproduce the empirical yields with explosion energies ranging from 3--10 B, even when a relevant SN explosion landscape is adopted (Section \ref{sec:compare_explosion}). %, suggesting this could be the explosion energy range that most CCSNe experience. 
However, when fixing explosion energy for a given progenitor mass with a functional form derived by \edit1{EC20} (i.e., imposing an explosion energy landscape) the theoretical yields disagree widely with the DLA observations. % in the mass range 10--100 M$_\odot$. 
LC18 (see Section \ref{sec:lc18_discuss}) reproduced most of the empirical yields with their 0 $\rm{km}~s^{-1}$ initial rotation velocity models, though some of the empirical yields were only reproduced by the higher rotation velocity models. %suggesting that this may not be an important parameter for the elements we compare in this work. 
When we apply a simple IDROV that mimics the bimodal velocity distribution of young massive stars in the local universe, there is near unanimous agreement between the predicted abundance ratios and empirical yields.  KN06 (Section \ref{sec:kn06_discuss}) adequately reproduced the empirical yields when contributions from HNe were $\gtrsim 50\%$ for most ratios, reinforcing the ability for high-energy explosions to match the DLA observations.

We found that the inclusion of realistic explosion physics (i.e., taking into account ranges of initial stellar mass that fail to explode) in the theoretical yields does not result in a quantitatively better fit to the empirical yields. \edit1{Interestingly, models from the PUSH collaboration or EC20 (see Section \ref{sec:pe20_discuss}), show the largest discrepancies in light element production (N) compared to the empirical yields, even though these models take into account the explosion landscape with initial stellar mass. These discrepancies likely originate in the pre-SN evolution of the star and not the modeling of the explosion (see Section \ref{sec:compare_explosion}).} This assertion is supported by \edit1{EC20} consistently reproducing the empirical ratios containing intermediate mass and heavy elements (e.g., Si, S, and Fe), which are more sensitive to the explosion mechanism, but not reproducing the ratios containing \edit1{N}.

Models that frequently reproduced the VMP DLAs abundance ratios had high explosion energies ($\geq$ 2 B\@). This finding adds more observational evidence to suggest that higher energies are helpful, perhaps necessary, in reproducing abundances measured from near pristine gas. However, these energies are believed to be unattainable with neutrino-driven explosions \citep[e.g.,][]{sukhbold+2016, ertl+2020}. There is some evidence that rotationally powered explosions, in which the progenitor gains rotational energy from its companion, could provide the necessary explosions energies \citep[][]{mosta+2015}. \edit1{However, the close match between the empirical yields and the predictions from EC20 of the intermediate mass elements (whose abundances are more sensitive to the explosion of the star) supports the opposite conclusion: HNe are not needed to explain  the abundance patterns if the explosion of the CCSN is modeled properly. A more detailed analysis of the Fe-peak element production between these different models could answer this question definitively.}  %Typically this is observed in metal poor stars, but is now being seen in VMP DLAs.

VMP DLAs have allowed for an empirically driven approach to quantify the abundances of metals ejected from the first stars. They are complementary to using VMP stars for the same purpose \citep[e.g.,][K19]{grimmett+2018,ishigaki+2018,griffith+2021a}. Improvements that would benefit future work using VMP DLAs include increasing the number of elemental abundances measured from VMP DLA spectra so that more abundance ratios can be constrained, and increasing the sample size of VMP DLAs to improve the statistics of the empirical constraints. Even with our current constraints (Table~\ref{tab:med_abundance_ratios}), an interesting avenue of inquiry would be to quantify how the results of galactic chemical evolution models \citep[e.g.,][de los Reyes (submitted)]{kirby+2011} change when theoretical input yields are replaced with our empirically estimated yields from VMP DLAs.

\section*{Acknowledgements}
We thank Tuguldur Sukhbold for their in-depth comments and correspondence that significantly improved this work. We thank Ryan Cooke, Louise Welsh, Donatella Romano, \edit1{Carl Fields}, Lynne Hillenbrand, and \edit1{Sanjana Curtis} for thoughtful conversations and for sharing information and data that improved the quality of this work. \edit1{We thank our referee for helpful comments that improved this paper.} E.H.N.\ acknowledges the support of the NSF Graduate Research Fellowship Program.  This material is based upon work supported by the National Science Foundation under Grant No.\ AST-1847909.  E.N.K.\ gratefully acknowledges support from a Cottrell Scholar award administered by the Research Corporation for Science Advancement.

\textbf{\textnormal{\software{Astropy \citep{astropy_2013, astropy_2018}}}}

\bibliography{citations}{}
\bibliographystyle{aasjournal}

\end{document}

%% file: med_vmpdlas_defs_v9_noESI.tex
\newcommand{\CFe}{0.16$\pm$0.20} 
\newcommand{\NFe}{-0.76$\pm$0.34} 
\newcommand{\OFe}{0.42$\pm$0.13} 
\newcommand{\AlFe}{-0.11$\pm$0.11} 
\newcommand{\SiFe}{0.32$\pm$0.16} 
\newcommand{\SFe}{0.36$\pm$0.18} 
\newcommand{\CO}{-0.30$\pm$0.13} 
\newcommand{\NO}{-1.21$\pm$0.33} 
\newcommand{\FeO}{-0.41$\pm$0.13} 
\newcommand{\AlO}{-0.55$\pm$0.16} 
\newcommand{\SiO}{-0.14$\pm$0.12} 
\newcommand{\SO}{-0.27}%$\pm$0.0} 